\documentclass[numbers=noenddot]{jfm}
\usepackage[latin9]{inputenc}

\usepackage{array, ragged2e}
\usepackage{float}
\usepackage{mathtools}
\usepackage{graphicx}
\usepackage{esint}
\usepackage{verbatim}
\usepackage{caption}
\usepackage{subcaption}
\renewcommand\[{\begin{equation}}
\renewcommand\]{\end{equation}}
\newcommand{\un}[1]{\,\mathrm{#1}}
\numberwithin{equation}{section}
\usepackage{amsmath} 
\newcommand{\beq}{\begin{equation}}
\newcommand{\eeq}{\end{equation}}
\newcommand{\lb}{\left(}
\newcommand{\rb}{\right)}
\usepackage{tikz}
\usepackage{pgfplots}
\usepackage{overpic}
\usepackage{hyperref}
\usepackage{todonotes}
\author{Graham P. Benham\aff{1,2}\corresp{\email{graham.benham@ladhyx.polytechnique.fr}},  
Jerome A. Neufeld\aff{1,2,3}, Andrew W. Woods\aff{1,2}}
\affiliation{\aff{1} Department of Earth Sciences, University of Cambridge, Cambridge CB3 0EZ, UK
\aff{2} Centre for Environmental and Industrial Flows, University of Cambridge, Cambridge CB3 0EZ, UK
\aff{3} Department of Applied Mathematics and Theoretical Physics, University of Cambridge, Cambridge CB3 0WA, UK
}

\begin{document}

\title{Axisymmetric gravity currents in anisotropic porous media}

\maketitle

\abstract{
We explore the motion of an axisymmetric gravity current in an anisotropic porous medium in which the horizontal permeability is larger than the vertical permeability. 
It is well known that the classical axisymmetric gravity current supplied by a constant point source of fluid has an unphysical singularity near the origin.
We address this by considering a pressure-dominated region near the origin which allows for vertical flow from the source, such that the current remains of finite depth, whilst beyond this region the flow is gravity-dominated. 
At early times the inner pressure-driven region controls the spreading of the current, but at late times the inner region occupies a progressively smaller fraction of the current such that the radius increases as $\sim t^{3/7}$, while the depth near the origin increases approximately as $\sim t^{1/7}$. 
The presence of anisotropy highlights this phenomenon, since the vertical permeability maintains an effect on the flow at late times through the pressure-driven flow near the origin. 
Using these results we provide some quantitative insights into the dominant dynamics which control  CO$_2$ migration through permeable aquifers, as occurs in the context of carbon capture and storage. }

\section{Introduction}

Buoyancy-driven flows in porous media resulting from the injection of fluid of different density to the original reservoir fluid have been studied in some detail owing to their importance for geothermal power production, carbon capture and storage and enhanced oil recovery, for example. The initial models of gravity-driven flow in a porous medium by \citet{barenblatt1996scaling} and \citet{huppert1995gravity} established some simple similarity solutions for two-dimensional gravity currents resulting from steady injection, and tested these models with analogue laboratory experiments in isotropic porous media. These models were based on the assumption that the vertical pressure gradient is everywhere hydrostatic and that the flow becomes long and thin, so that the predominantly horizontal flow is driven by the horizontal pressure gradient associated with variations in the thickness of the current. The solution takes the form
\beq
h/L = ({t}/{\tau})^{1/3} f\left[(x/L)(t/\tau)^{-2/3}\right],
\eeq
where $L=Q_x\mu/k\Delta \rho g$ and $\tau=Q_x(\mu/k\Delta \rho g)^2$ are dimensional length and time scalings, given in terms of the input flow rate per unit length, $Q_x$, the permeability, $k$ (assumed isotropic), the density difference between the two fluids, $\Delta\rho$, and the viscosity of the injected fluid, $\mu$ \citep{huppert1995gravity}.
If one extends the analysis to account for different permeabilities in the horizontal and vertical directions, $k_H$ and $k_V$, the assumption of hydrostatic pressure in the flow means that the vertical permeability does not feature in the solution, which is the same as for the isotropic case. 
This is curious since the vertical permeability of the porous medium may be much smaller than the horizontal permeability owing to the original geological processes of formation and compaction of the medium \citep{corbett1992variation,martinius1999multi,woods2015flow}. The key to this apparent paradox is to assess the vertical pressure gradient required to drive the vertical flow which is implicit in the gravity current solution. With a very small vertical permeability, this pressure gradient scales as
 \beq
\frac{\partial p}{\partial z}\sim  \frac{\mu  } { k_V}\frac{\partial h}{\partial t} \sim \frac{\mu  L}{\tau^{1/3}t^{2/3}  k_V}.\label{introeq}
 \eeq
For the solution to be valid we require that \eqref{introeq} is small compared to the corresponding hydrostatic pressure gradient
 \beq
\frac{\partial p}{\partial z}\sim { \Delta \rho g }.
 \eeq
This requires that
 \beq
 t/\tau\gg \lb{k_H}/{k_V}\rb^{3/2},
 \eeq
 and we see the role of the vertical permeability in determining the time at which the flow adjusts to the similarity solution. It is worth noting that this result provides an extension of the analysis presented by \citet{huppert2022fate} who considered the early time pressure solution in isotropic media. 
 
 For radial injection from a point source, the situation is more complex. In the classical similarity solution \citep{lyle2005axisymmetric} there is a singularity in the calculated depth of the current at the origin. However, this result is unphysical, and in order that the flow remains of finite depth at the origin there is a near-origin adjustment zone for all time. Inclusion of this near-origin adjustment zone leads to a different set of scaling laws and a new solution for the current that is influenced by $k_V$ for all time. 
In this study we address these details using a theoretical analysis in conjunction with numerical simulations, and we show the analysis is consistent with a series of new laboratory experiments. Our study goes beyond that of \citet{huppert2022fate}, where it was assumed that the pressure-driven flow simply adjusts to the classical similarity solution.

We consider the application of our results to the case of CO$_2$ sequestration and illustrate how  anisotropy of the porous medium leads to the ratio $k_V/k_H$ influencing the dispersal of the CO$_2$ at all times, in contrast to the classical gravity current solution. 
Furthermore, we show that for typical injection rates the transition in flow regime from a pressure-dominated flow to a gravity-dominated flow may occur once the flow has reached depths of $10-1000$ m at time scales of $1-10$ years, depending on the value of $k_V/k_H$. This suggests that the initial pressure-driven flow persists for a significant period of the injection, and in some cases the flow may never become gravity-driven over the time and length scales relevant to those sites.


The structure of the paper is laid out as follows. Section 2 addresses the flow scenario, treated analytically at early and late times, and comparisons are made with both numerical simulations and porous bead experiments. Section 3 applies the results to the context of carbon sequestration, calculating the criteria for whether several sites are in a pressure- or gravity-driven regime, and Section 4 closes with a discussion of these results.

\section{Axisymmetric injection into anisotropic porous media}

\subsection{A note on anisotropic permeability}

Before discussing the flow scenario, we first briefly discuss the geological context of anisotropy in porous media.
In subsurface geological reservoirs it is common for the permeability of rocks to differ significantly depending on the direction of the flow \citep{corbett1992variation,martinius1999multi}. 
This may result from post-depositional compaction of the formation or from the deposition of successive layers of fine and coarse material. 
Hence, the permeability field is sometimes written as a three-dimensional diagonal matrix, $\mathbf{k}=\mathrm{Diag}(k_x,k_y,k_z)$, or equivalently in a different basis depending on the direction of compaction (e.g. see studies on cross-bedding \citep{woods2015flow}).

We account for such situations, but we restrict our attention to the case where the compaction is aligned with the vertical coordinate, and hence the permeability is reduced to horizontal and vertical variation only, $(k_x,k_y,k_z)=(k_H,k_H,k_V)$, where $k_H$ and $k_V$ are constants. Hence, the anisotropy is characterised by the single dimensionless parameter 
\beq
K=k_H/k_V.
\eeq
This formulation is relevant either for flow through a single thick sedimentary layer that has been compacted vertically, resulting in a binary permeability field, or for flow through a system of many horizontal sedimentary layers, in which the values $k_H$, $k_V$, can be interpreted as \textit{effective} permeabilities \citep{cardwell1945average,kumar1997efficient,woods2015flow}. For example, in the latter scenario, if the layers vary between permeabilities $k_1$ and $k_2$ over alternating layer widths $d_1$ and $d_2$, then the effective permeability values are given by the arithmetic and harmonic mean values
\beq
k_H\approx \frac{d_1 k_1 +d_2k_2}{d_1+d_2} ,\quad k_V\approx \frac{d_1+d_2}{d_1/k_1+d_2/ k_2}.  \label{permapprox}
\eeq
This tends to be a good approximation as long as the depth of the flow is much larger than the widths of the layers $d_1,d_2$. 
Hence, in this system of horizontal sedimentary layers the anisotropy is always such that $K\geq 1$.
In fact, for many geological systems the horizontal and vertical permeability have been observed to differ by several orders of magnitude \citep{martinius1999multi}, resulting in anisotropy values as large as $K=\mathcal{O}(10^4)$.

\subsection{The flow scenario at early times and subsequent transition behaviour}

Next we move on to describe the flow scenario we consider, and describe how this evolves at early times, before the effects of gravity become significant. 
For the isotropic case, as discussed by \citet{huppert2022fate}, the constant input of fluid from a point source results in the current expanding in the shape of a self-similar hemisphere at early times. The radial and vertical extent of the current are equal and grow like the cube root of time. 
The flow transitions to a gravity-driven regime once the weight of the fluid dominates over the injection pressure.
In the following analysis, we extend this early-time analysis to the case of injection into an anisotropic medium, demonstrating how the anisotropy $K$ affects both the early-time dynamics and the transition behaviour.

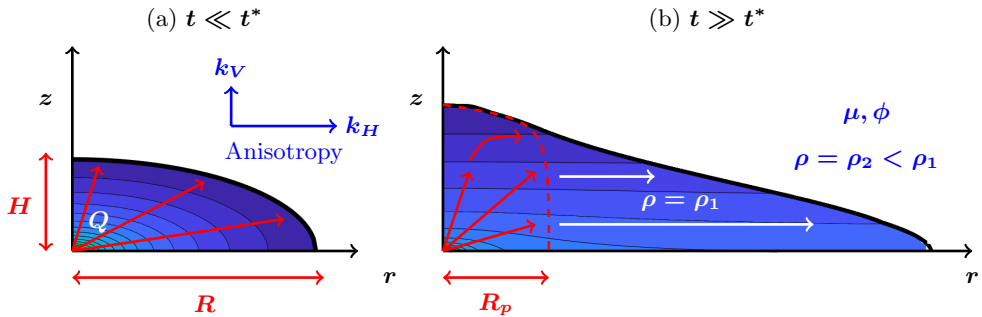
\begin{figure}
\centering
\begin{tikzpicture}[scale=0.7]
\node at (12.4,4.3) {\includegraphics[width=0.616\textwidth]{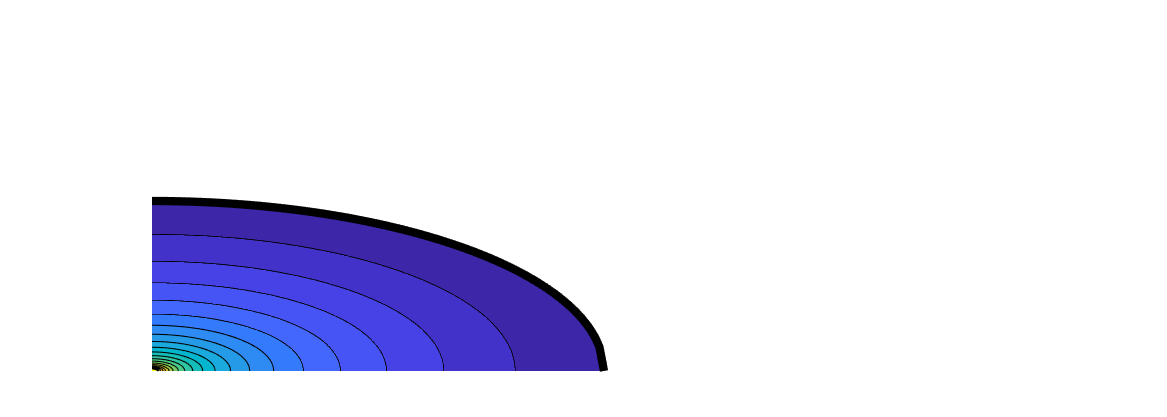}};
\draw[line width=1, black,->]  (8,2.65) -- (13.4,2.65);
\draw[line width=1, black,->]  (8,2.65) -- (8,6.5);
\node at (7.5,5.5) {$\boldsymbol{z}$};
\draw[line width=1, red,<->]  (7.5,2.65) -- (7.5,4.5);
\draw[line width=1, red,<->]  (8,2.15) -- (12.75,2.15);
\node[red] at (7,3.5) { $\boldsymbol{H}$};
\node[red] at (10.5,1.65) { $\boldsymbol{R}$};
\draw[line width=1, red,->]  (8,2.65) -- (10.5,3.95);
\draw[line width=1, red,->]  (8,2.65) -- (12.0,3.25);
\draw[line width=1, red,->]  (8,2.65) -- (8.5,4.25);
\node[white] at (8.5,3.2) { \rotatebox{0}{$\boldsymbol{Q}$}};
\node at (19.4,4.3) {\includegraphics[width=0.616\textwidth]{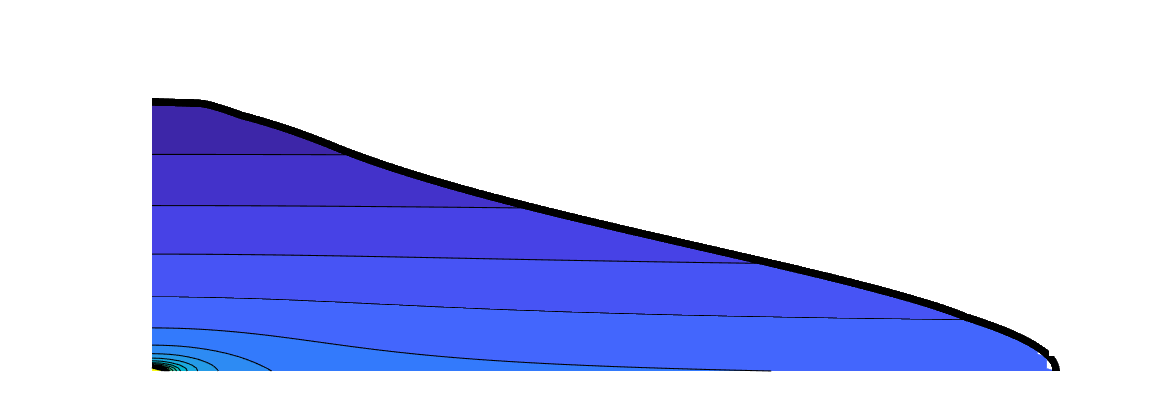}};
\draw[line width=1, black,->]  (15,2.65) -- (25,2.65);
\draw[line width=1, black,->]  (15,2.65) -- (15,6.5);
\node at (14.5,5.5) {$\boldsymbol{z}$};
\node at (25,2.15) {$\boldsymbol{r}$};
\node at (14,2.15) {$\boldsymbol{r}$};
\draw[line width=1, red,->]  (15,2.65) -- (16.75,3.15);
\draw[line width=1, red,->]  (15,2.65) -- (16.75,4.15);
\draw[line width=1, red,->]  (15,2.65) -- (15 .5,4.15);
\draw[line width=1,<->,red]  (15,2.15) -- (17,2.15);
\node[red] at (16,1.65) { $\boldsymbol{R_p}$};
\draw[line width=1, red,dashed]  (15,5.4) .. controls (17,5.1) .. (17,2.65);
\draw[line width=1, red,->]  (15.5,4.35) .. controls (15.8,4.8) .. (16.5,4.8);
\draw[line width=1, white,->]  (17.2,4.05) -- (19,4.05);
\draw[line width=1, white,->]  (17.2,3.15) -- (22,3.15);
\node[white] at (19.5,3.55) { \bf $\boldsymbol{\rho=\rho_1}$};
\node[blue] at (23,4.35) { \bf $\boldsymbol{\rho=\rho_2<\rho_1}$};
\node[blue] at (23,5.25) { \bf $\boldsymbol{\mu,\phi}$};
\node[blue] at (11,6.1) { $\boldsymbol{k_V}$};
\node[blue] at (13.5,5) { $\boldsymbol{k_H}$};
\node[blue] at (12,4.5) { Anisotropy};
\draw[line width=1, blue,->]  (11,5) -- (13,5);
\draw[line width=1, blue,->]  (11,5) -- (11,5.75);
\node at (20,7) {(b) \bf $\boldsymbol{t\gg t^*}$};
\node at (10.5,7) {(a) \bf $\boldsymbol{t\ll t^*}$};
\end{tikzpicture}
\caption{Schematic diagram illustrating the transition from a pressure-driven flow at early times (a) to a gravity-driven flow with a pressure-driven boundary layer near the origin at late times (b) in an anisotropic porous medium. 
Contour lines are indicative of the typical pressure field and approximate streamlines are sketched. 
\label{schem1}}
\end{figure}

We consider the constant axisymmetric injection of fluid (at flow rate $Q$) into a porous medium with anisotropy $K$ and porosity $\phi$, bounded from below by an impermeable substrate, as illustrated in figure \ref{schem1}. The porous medium is initially saturated with an ambient phase which has a lighter density than the injected fluid $\rho_2<\rho_1$, whereas the viscosities of the two fluids are assumed to be equal $\mu_1=\mu_2=\mu$. For simplicity, we assume that there is no mixing between the these two fluids, such that the interface between them remains sharp. 
We choose a cylindrical polar coordinate system $(r\geq0,z\geq0)$ and we denote the radial and vertical extent of the current, $R(t)$ and $H(t)$, which increase over time due to the constant injection rate $Q$ from a point at the origin.

At early times the pressure is dominated by the viscous resistance due to injection and gravity has a negligible effect. Hence, due to conservation of mass the pressure must satisfy Laplace's equation. However, since the medium is anisotropic this results in non-uniform coefficients, such that
\beq
k_H\frac{1}{r}\frac{\partial}{\partial r} \lb r \frac{\partial p}{\partial r} \rb +k_V\frac{\partial^2 p}{\partial z^2}=0.
\eeq
Since $k_H$ and $k_V$ are constants, the dynamics can be easily mapped from an isotropic medium, and hence the lateral and vertical extents must satisfy
\beq
H^2/R^2=1/K.\label{laplace}
\eeq
Likewise, conservation of mass (within a stretched hemisphere) dictates that
\beq
2\pi \phi HR^2/3 = Q t.\label{earmassax}
\eeq
This indicates that the early time dynamics are described by
\begin{align}
H&=K^{-1/3}(3Qt/2\pi\phi)^{1/3},\label{early_2d_anis1}\\
R&=K^{1/6}(3Qt/2\pi\phi)^{1/3}.\label{early_2d_anis2}
\end{align}
Hence, at early times the effect of anisotropy is to stretch/squash the flow in the more/less permeable direction.
It follows that the thickness of the current, $z=h(r,t)$, is given by
\beq
h/H=\left[ 1-(r/R)^2\right]^{1/2},\label{isotropic}
\eeq
at early times (see figure \ref{twod_stream}a).
It is surprising to note that (within this early time regime) a strongly anisotropic medium results in a current which is very long and thin, but in which the pressure is not hydrostatic. This is contrary to the common assumption that long and thin currents automatically imply a hydrostatic pressure.


After a significant amount of time the current grows to a thickness where the weight of fluid begins to dominate over the injection pressures. This is equivalent to the moment when the input flow rate $Q$ is matched by the flow rate due to hydrostatic pressure gradients. For a hydrostatic current, we have that $\partial p/\partial r\approx \Delta \rho g \partial h/\partial r$, so that the integrated gravity-driven flux scales like
\beq
Q_g\sim-\frac{k_H\Delta \rho g}{\mu} r h\frac{\partial h}{\partial r}\sim Q.\label{grav_balance0}
\eeq
By inserting approximate scalings, $h\sim H$, $r\sim R$, $\partial h/\partial r\sim -H/R$, this provides the balance
\beq
Q_g\sim  u_b  H^2 \sim Q,\label{grav_balance}
\eeq
where the buoyancy velocity $u_b=k_H\Delta \rho g/\mu$. This immediately provides an expression for the thickness of the current at the transition between pressure- to buoyancy-driven flow,
\beq
H^*=(Q/u_b)^{1/2},\label{times_radial0}
\eeq
and by equating this to the early time behaviour \eqref{early_2d_anis1} we find that the transition time is
\beq
t^*= (2K\phi\pi/3)(Q/ u_b^3)^{1/2}.\label{transtimescale}
\eeq
Therefore, anisotropy causes the transition to occur at later times, as expected. Indeed, the significance of this delayed transition is immediately appreciable when one considers that some geological formations have anisotropy values as large as $K=\mathcal{O}(10^4)$ \citep{martinius1999multi,bergmo2017quality}.

The critical time $t^*$ distinguishes two distinct regimes. At early times $t\ll t^*$, or when $H\ll H^*$, the effects of gravity are negligible and injection pressures dominate the flow, whereas at late times $t\gg t^*$, or when $ H\gg H^*$, gravity plays a significant role. However, clearly the flow must remain pressure-driven very close to the source, even at late times. This pressure-driven region is responsible for distributing the flow across the vertical extent of the current, whereupon it diverts to the remaining gravity-driven region far away from the origin (see figure \ref{schem1}b). Whilst various authors have alluded to the existence of this pressure-driven boundary layer, it is not known how its lateral and vertical extent evolve over time, nor how it couples with the remaining gravity current. Furthermore, it is unknown how anisotropy affects the flow after transition occurs, which is the subject of the following sections.

\subsection{Late-time dynamics: classical analysis}

At much later times $t\gg t^*$, once the flow has grown to a thickness $H\gg H^*$, it retains much of the character of a classical porous gravity current. \citet{lyle2005axisymmetric} described the late-time dynamics in an isotropic porous medium by assuming a hydrostatic pressure profile everywhere within the current. This results in the corresponding radially symmetric thin-film equation
\beq
\phi\frac{\partial h}{\partial t}=u_b\frac{1}{r}\frac{\partial }{\partial r}\left[r h \frac{\partial h}{\partial r}\right],\label{thinfilm}
\eeq
for the evolution of the current thickness $z=h(r,t)$. The above equation is accompanied by boundary conditions imposing zero thickness at the nose, constant injection flux at the origin, and zero flux through the nose, such that
\begin{align}
h&=0,\quad \mathrm{at}\quad r=R(t),\label{zeronose}\\
-2\pi r h \frac{\partial h}{\partial r}&\rightarrow Q, \quad \mathrm{as}\quad r\rightarrow 0,\label{dimbc1}\\
-2\pi r h \frac{\partial h}{\partial r}&\rightarrow 0, \quad \mathrm{as}\quad r\rightarrow R(t).\label{dimbc2}
\end{align}
It should be noted that, whilst these equations were formulated for the case of an isotropic porous medium, the analysis can be easily transposed to the case of anisotropic permeability, $k_H$, $k_V$. However, the vertical permeability $k_V$ vanishes from the analysis since the injected flow and the ambient flow are decoupled (an assumption which we revisit later), resulting in an identical system of equations.

\begin{figure}
\centering
\begin{tikzpicture}[scale=1.25]
\node at (0,-4) {\includegraphics[width=0.4\textwidth]{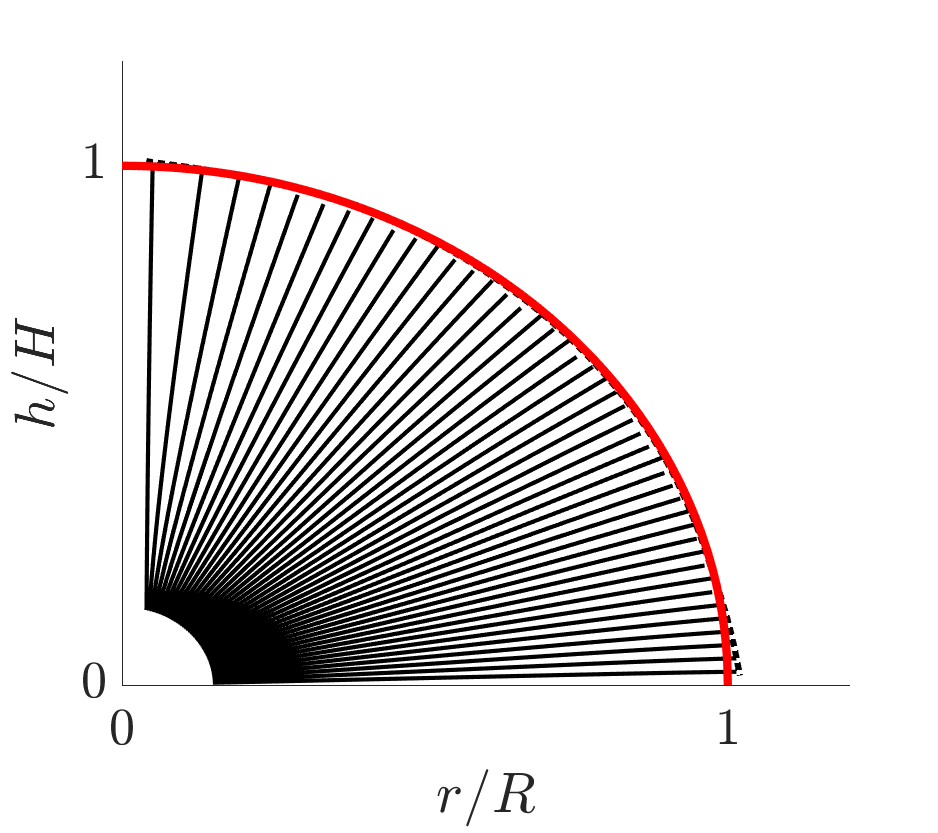}};
\node at (3.1,-2.9) {\includegraphics[width=0.25\textwidth]{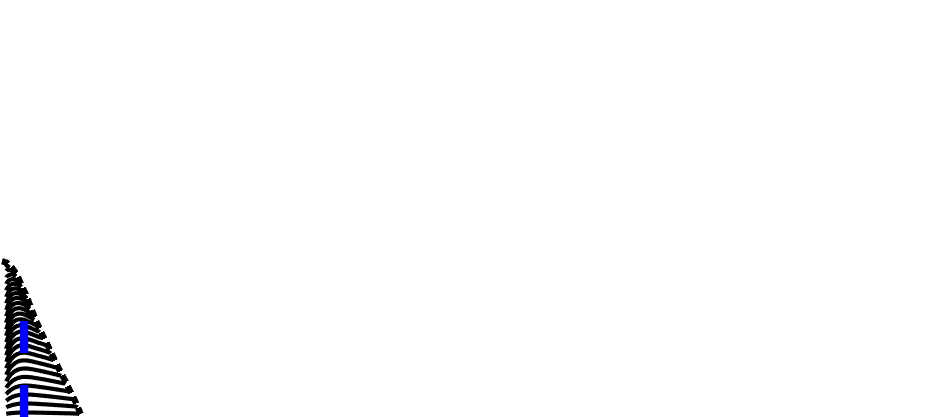}};
\node at (4.2,-2.9) {\includegraphics[width=0.25\textwidth]{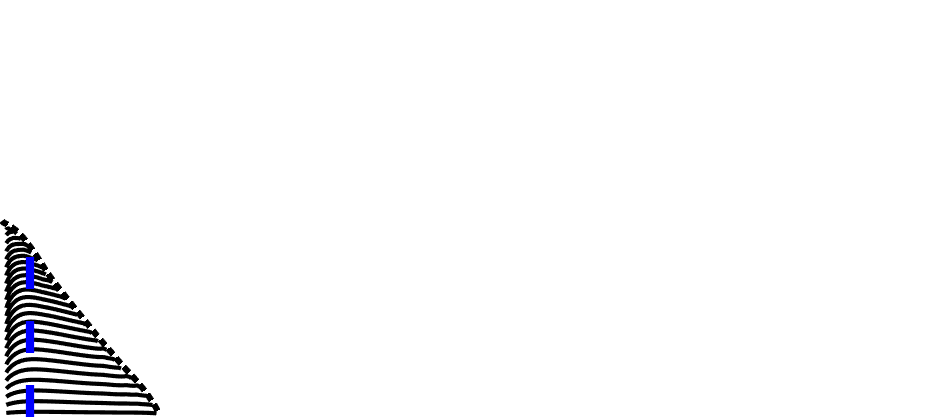}};
\node at (3.1,-4.2) {\includegraphics[width=0.25\textwidth]{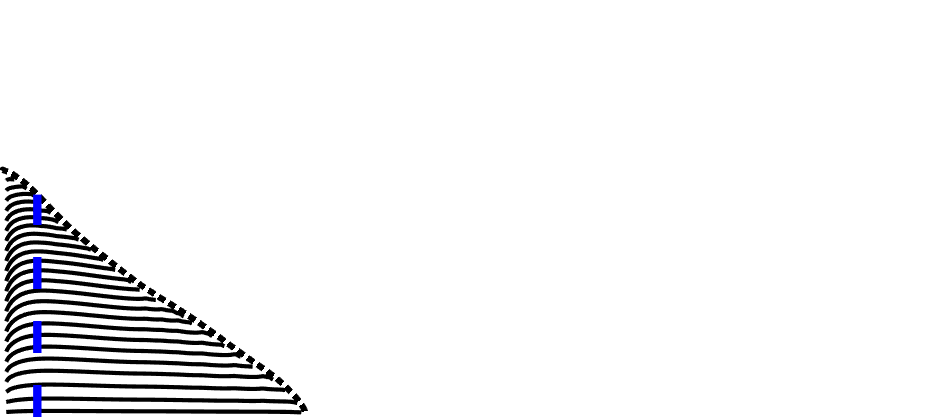}};
\node at (4.2,-4.2) {\includegraphics[width=0.25\textwidth]{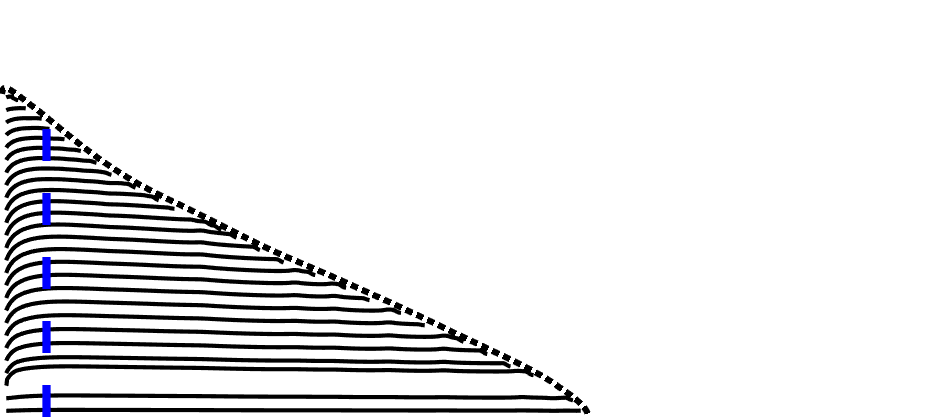}};
\node at (6.7,-4) {\includegraphics[width=0.4\textwidth]{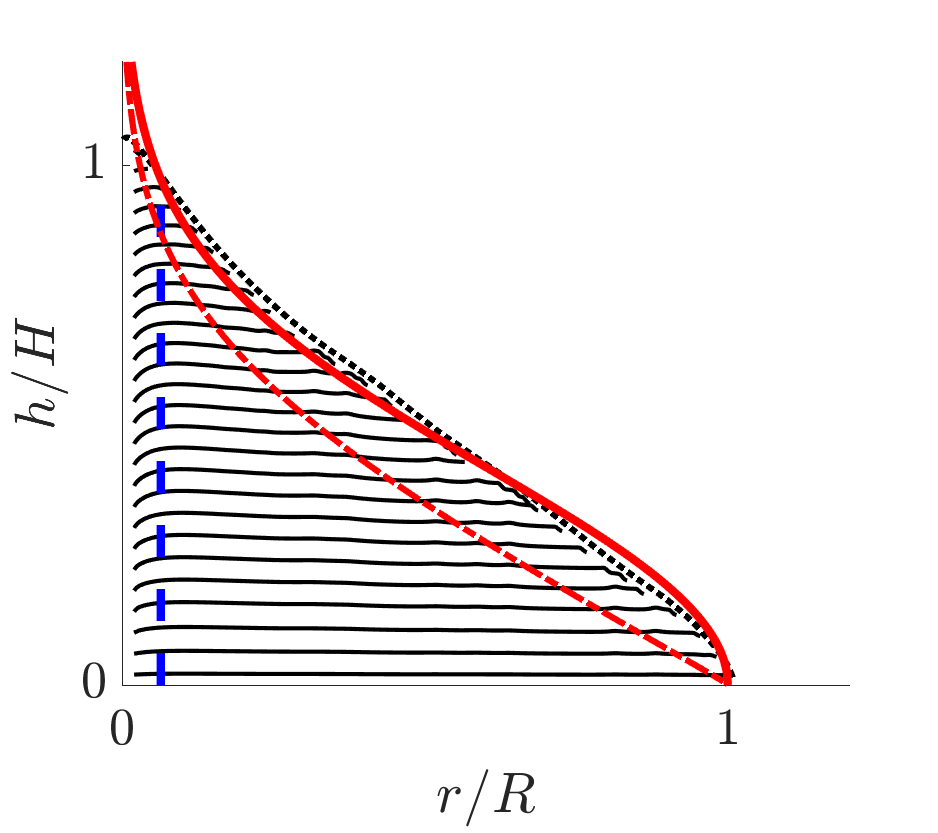}};
\node[blue] at (5.9,-2.8) {\bf $\boldsymbol{R_p(t)}$};
\draw[line width=1.5, blue,->]   (5.6,-4.25) --  (5.3,-4.25);
\draw[line width=1.5, blue,->]   (5.6,-4.65) --  (5.3,-4.65);
\draw[line width=1.5, blue,->]   (5.6,-5.05) --  (5.3,-5.05);
\node at (-1.9,-2.5) {(a)};
\node at (2,-2.5) {(b)};
\node at (4.6,-2.5) {(c)};
\end{tikzpicture}
\caption{Numerical solution of the current shape $z=h(r,t)$, illustrating analytical solutions with red lines.  Early-time profile in (a) ($t/t^*=10^{-3}$) is normalised by early scalings \eqref{early_2d_anis1}-\eqref{early_2d_anis2}. Transition regime profiles (b) ($t/t^*=1-10^2$) have no normalisation. Late-time profile in (c) ($t/t^*=10^3$) is normalised by late scalings \eqref{late_rad_isoH}-\eqref{late_rad_isoR}.  The position of the late-time boundary layer near the origin $r=R_p(t)$ (relatively shrinking) is indicated in blue. \label{twod_stream}}
\end{figure}
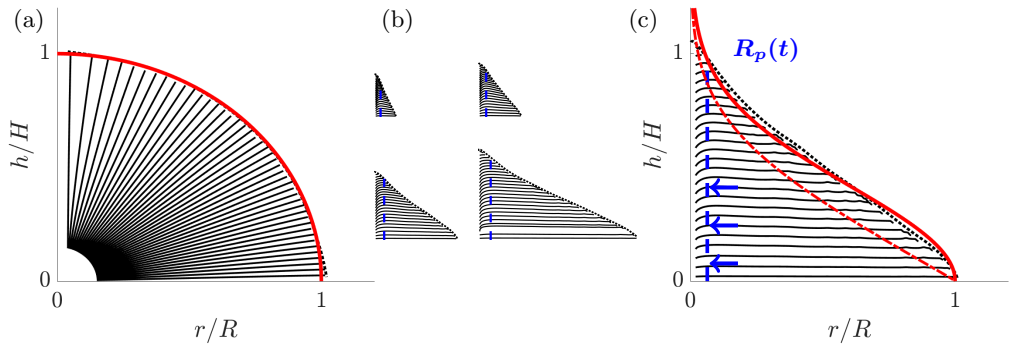

The system of equations \eqref{thinfilm}-\eqref{dimbc2} admits the similarity variables
\beq
h=(Q/u_b)^{1/2} f(\eta/\eta_N),\quad \eta =r(Q u_b)^{-1/4}(t/\phi)^{-1/2},\label{simvars}
\eeq
where the dimensionless shape $f$ and prefactor $\eta_N$ satisfy
\begin{align}
-\frac{1}{2}\eta f'&=\frac{1}{\eta}\frac{\mathrm{d}}{\mathrm{d}\eta}\left[ \eta f \frac{\mathrm{d} f}{\mathrm{d}\eta}\right],\label{govsim}\\
f&=0, \quad \mathrm{at}\quad \eta = \eta_N,\label{zeroh}\\
-2\pi \eta f \frac{\mathrm{d} f}{\mathrm{d}\eta} & \rightarrow 1, \quad \mathrm{as}\quad \eta \rightarrow 0,\label{influx}\\
-2\pi \eta f \frac{\mathrm{d} f}{\mathrm{d}\eta} & \rightarrow 0, \quad \mathrm{as}\quad \eta \rightarrow \eta_N. \label{outflux}
\end{align}
The solution can be calculated numerically, giving a prefactor value $\eta_N=1.155$, and the resultant shape is plotted in figure \ref{twod_stream}c as a red dash-dot line. The influx condition \eqref{influx} results in singular behaviour of the thickness $h$ near the origin, which is a consequence of the hydrostatic pressure assumption.
This is unphysical and can be addressed by considering the pressure-driven flow near the origin.
In particular, the flux within this region is not buoyancy-driven, and so the inflow boundary condition \eqref{influx} (resulting in singular behaviour of the thickness) no longer applies. Instead, the inflow condition requires a non-hydrostatic flux component within the pressure-driven zone, and as we will see, this maintains a dominant contribution to the flow (even at late times) such that the thin-film approximation \eqref{thinfilm} cannot be used near the origin.

\subsection{Back to the future: late-time dynamics revisited}\label{sec_seventh}

\begin{figure}
\centering
\begin{tikzpicture}[scale=1]
\node at (19.4,4.3) {\includegraphics[width=0.88\textwidth]{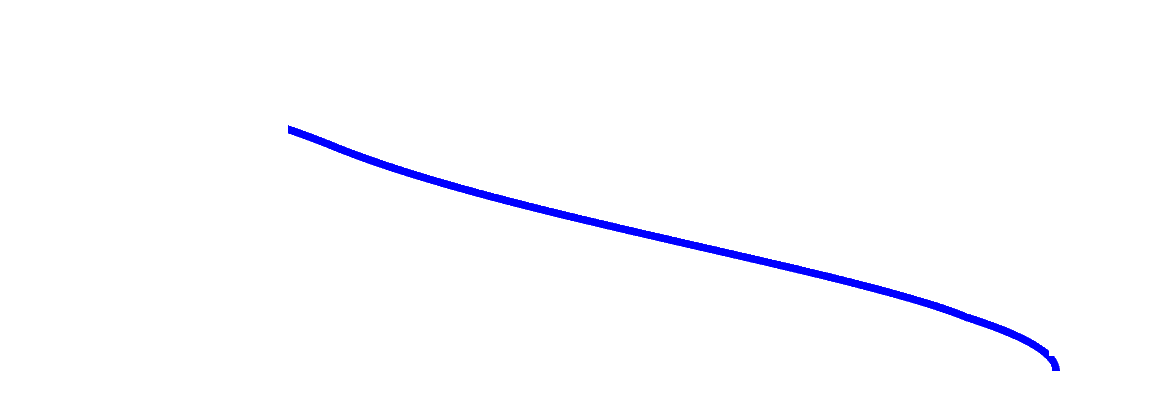}};
\draw[line width=1, black,->]  (15,2.65) -- (25,2.65);
\draw[line width=1, black,->]  (15,2.65) -- (15,6.5);
\node at (14.5,6.2) {$\boldsymbol{z}$};
\node at (25,2.15) {$\boldsymbol{r}$};
\draw[line width=1,<->,red]  (15,2.15) -- (17,2.15);
\draw[line width=1,<->,blue]  (15,2.45) -- (24.2,2.45);
\draw[line width=1,<->,red]  (14.7,2.65) -- (14.7,5.4);
\node[red] at (16,1.65) { $\boldsymbol{R_p\sim\xi_0\, t^{a}}$};
\node[blue] at (20,1.85) { $\boldsymbol{R\sim\zeta_N \,t^{b}}$};
\node[red] at (14.2,4.05) { \rotatebox{90}{$\boldsymbol{H\sim\xi_0\, t^{a}}$}};
\draw[line width=2, red]  (15,5.4) .. controls (17,5.1) .. (17,2.65);
\node[black] at (16,4.25) {\bf Inner};
\node[black] at (16,3.65) {\bf (pressure-};
\node[black] at (16,3.25) {\bf driven)};
\node[black] at (20,3.6) {\bf Outer};
\node[black] at (20,3.) {\bf (gravity-driven)};
\end{tikzpicture}
\caption{Schematic diagram showing the different variables of the late-time regime. The analysis in Section \ref{sec_seventh} shows that the power law values are $a=1/7,b=3/7$. Note the aspect ratio is exaggerated for illustration purposes, but in fact $H$ is smaller than $R_p$ due to anisotropy \eqref{laplace2}. 
\label{schem2}}
\end{figure}
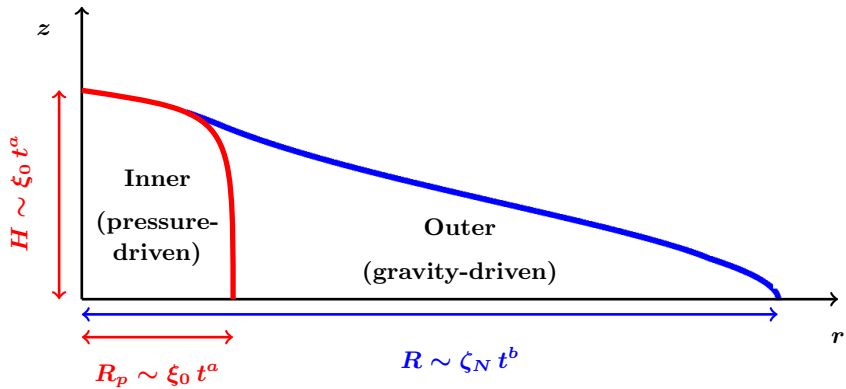

Within the inner pressure-driven region strong vertical velocities deliver the injected flow all the way to the uppermost extent of the current.  
The inner pressure-driven radius and the thickness scaling are related (for the same reasons as \eqref{laplace}) according to 
\beq
H^2/R_p^2=1/K.\label{laplace2}
\eeq
Our isotropic porous bead experiments, discussed later in Section \ref{sec_exp}, show that the thickness of the current near the origin increases slowly over time (e.g. see figure \ref{expt3}). 
Therefore, we attempt to describe the solution within this inner region using a set of rescaled variables where we allow for the possibility of an evolving thickness at the origin.
Using a weak power law value  $0<a<1$, the variables are rescaled according to
\beq
\xi=r/R_p(t),\quad h=K^{-1/2}\mathcal{G}(\xi,t)R_p(t),\quad R_p(t)=\xi_0 u_b^{\frac{3a-1}{2}}Q^{\frac{1-a}{2}} (t/\phi)^a,\label{rescale1}
\eeq
where $\mathcal{G}$ is the dimensionless shape function, $\xi_0$ is a dimensionless constant that we determine later, and the powers of $u_b$ and $Q$ in \eqref{rescale1} are chosen to be dimensionally consistent.

The consequence of having a growing inner region is that the radial extent of the corresponding {outer} region does not grow according to the classical $R\sim t^{1/2}$ power law described earlier. In particular, conservation of mass (in conjunction with \eqref{laplace2}) dictates that
\beq
\pi\phi  H R^2\sim \pi\phi  K^{-1/2}R_p R^2\sim Q t.\label{massconlate}
\eeq
If $a>0$, it follows that $R$ grows as $\sim t^b$, where $b<1/2$.
Hence, we introduce a corresponding set of rescaled outer variables
\beq
\zeta=r/R(t),\quad h=K^{-1/2}\mathcal{F}(\zeta,t)R_p(t),\quad R(t)=\zeta_N u_b^{\frac{3b-1}{2}}Q^{\frac{1-b}{2}}  (t/\phi)^b,\label{rescale2}
\eeq
where $\mathcal{F}$ is the dimensionless shape profile, the nose position $\zeta_N$ is a dimensionless constant to be determined, and $h$ is rescaled to match with the inner region. Mass conservation \eqref{massconlate} indicates that
\beq
a+2b=1.\label{masscoeff}
\eeq
Given these rescaled variables \eqref{rescale2}, the thin film equation \eqref{thinfilm} in the outer (hydrostatic) region becomes
\beq
\frac{R^2}{R_p}\frac{\partial \mathcal{F}}{\partial t}-\frac{R R'}{R_p}\zeta\frac{\partial \mathcal{F}}{\partial \zeta}+\frac{R^2 R_p'}{R_p^2}\mathcal{F}=\frac{u_b}{ K^{1/2}\phi\zeta}\frac{\partial }{\partial \zeta}\left[\zeta \mathcal{F} \frac{\partial \mathcal{F}}{\partial \zeta}\right].\label{thinfilm2}
\eeq
The second two terms on the left hand side of \eqref{thinfilm2} are of the order $\mathcal{O}(t^{-2a})$ and so become vanishingly small at large times. Assuming the time derivative of the scaled shape decreases as ${\partial \mathcal{F}}/{\partial t}\sim1/t$ for large times, then the first term on the left hand side is also of the order $\mathcal{O}(t^{-2a})$. Hence, after sufficiently late times \eqref{thinfilm2} yields the result
\beq
-2\pi \zeta \mathcal{F} \frac{\mathrm{d} \mathcal{F}}{\mathrm{d} \zeta}\approx \mathcal{Q},\label{radflux}
\eeq
where $\mathcal{Q}$ is the scaled flux of injected fluid, which is uniform in magnitude across the extent of the gravity current, and which is yet unknown. Upon further integration and applying the boundary condition \eqref{zeronose}, we arrive at an expression for the outer solution, which is
\beq
\mathcal{F}\approx\lb -\frac{\mathcal{Q}}{\pi} \log \zeta \rb^{1/2}.\label{approxsol}
\eeq
We note that the integrated gravity-driven flux, which is given by 
\beq
Q_g=-2\pi u_b {r} {h} \frac{\partial {h}}{\partial {r}},\label{gravfluxdef}
\eeq
grows according to
\beq
Q_g=K^{-1}\mathcal{Q}u_b R_p(t)^2.\label{pre_flux}
\eeq
At the transition thickness, $H=K^{-1/2}R_p=H^*=(Q/u_b)^{1/2}$, the gravity-driven flux \eqref{pre_flux} equals the input flux $Q$, indicating that $\mathcal{Q}=1$.
We also note that, after the transition time $t>t^*$, the gravity-driven flux \eqref{pre_flux} continues to grow unbounded over time. Hence, this can satisfy neither the boundary condition at the origin \eqref{dimbc1}, nor that at the nose \eqref{dimbc2}. Losing the ability to impose the boundary condition at the nose indicates that a shock profile will form, over which the flux discontinuously drops to zero (this is a feature of such equations, e.g. see \citep{whitham2011linear})  Hence, in terms of the classification of partial differential equations, the problem is hyperbolic, unlike the case of a two-dimensional gravity current due to a line source \citep{huppert1995gravity} which is parabolic.

Next we consider conservation of mass to determine the model parameters. In particular, \eqref{thinfilm}, \eqref{dimbc1} and \eqref{dimbc2} can be combined to give a single volume integral
\beq
2\pi\phi \int_0^R r h \, \mathrm{d}r =Q t.\label{massint}
\eeq
Hence, by inserting \eqref{approxsol} into \eqref{massint}, we arrive at a relationship involving $\zeta_N$ and $\xi_0$, which is
\beq
2\pi \zeta_N^2\xi_0 (\pi K)^{-1/2} \int_{\epsilon(t)}^{1} y (-\log y)^{1/2} \, \mathrm{d}y = 1,\label{massrel1}
\eeq
where $\epsilon=R_p(t)/R(t)=\mathcal{O}(t^{-(b-a)})$. Noting that $\epsilon$ becomes vanishingly small at late times (since $b>a$), and noting that $\int_0^1 y (-\log y)^{1/2} \, \mathrm{d}y =(\pi/32)^{1/2}$, we can rearrange \eqref{massrel1} to give
\beq
 \zeta_N^2\xi_0 \approx {{(8K)^{1/2}}}/{\pi}.\label{coeff1}
\eeq
A further final condition is required to fully determine the coefficients $\zeta_N$, $\xi_0$, and this is provided by matching the flux of the current between the inner and outer regions. 
Moving from the outer region to the origin the thickness tends to a finite but growing value $h=H(t)$ (with zero slope) and a balance is sustained at the matching radius $r=R_p$, so the gravity-driven flux becomes
\beq
-2\pi u_b {r} {h} \frac{\partial {h}}{\partial {r}}= 2\pi K^{-1} u_b {R_p^3}{R^{-1}}= Q,\quad\mathrm{at}\quad r=R_p(t).\label{flux_match}
\eeq
Hence, \eqref{flux_match} suggests that
\beq
b=3a,\label{fluxcoeffs}
\eeq
which together with \eqref{masscoeff}, gives the power law values
\beq
a=1/7, \quad b=3/7.\label{powlawvals}
\eeq 
Furthermore, \eqref{flux_match} implies that the coefficients satisfy
\beq
 \xi_0^3/\zeta_N = K/2\pi.\label{coeff3}
\eeq
Hence, the two equations \eqref{coeff1} and \eqref{coeff3}, can be solved simultaneously to give
\begin{align}
\xi_0=(2^{1/2}\pi^3)^{-1/7}K^{5/14}&\approx 0.58K^{5/14},\\
\zeta_N=(2^{11/2}\pi^{-2})^{1/7}K^{1/14}& \approx 1.24K^{1/14}.
\end{align} 
In addition, the thickness and radial scalings are given by
\begin{align}
H&=K^{-1/7}(Q^3 u_b^{-2}\pi^{-3}2^{-1/2})^{1/7} (t/\phi)^{1/7},\label{late_rad_isoH}\\
R&=K^{1/14}(2^{11/2}Q^2u_b \pi^{-2})^{1/7} (t/\phi)^{3/7},\label{late_rad_isoR}\\
R_p&=K^{5/14}(Q^3 u_b^{-2}\pi^{-3}2^{-1/2})^{1/7} (t/\phi)^{1/7},\label{late_rad_isoRp}
\end{align}
which now all reflect a dependence on the anisotropy $K$. 
We note one further detail of this analysis. Whilst $H(t)$ \eqref{late_rad_isoH} captures the overall thickness scaling of the gravity-current, it does not reflect the precise value at the origin. In particular, if we insert the matching radius $r=R_p$ into the thickness profile in the outer region \eqref{approxsol}, we find that the thickness near the origin actually grows like
\beq
h(R_p(t),t)= K^{-1/2}R_p(-(1/\pi) \log R_p/R)^{1/2}\sim (2/(7\pi K) \log t)^{1/2} t^{1/7}. \label{logterm}
\eeq
However, since the relative growth of the logarithmic component is very slow, this only becomes appreciable at time scales $t/t^*$ larger than around $e^{7\pi K/2}\approx (5.96\times 10^{4})^K$. To summarise, the scaling for $H(t)$ \eqref{late_rad_isoH} provides the general behaviour of the thickness everywhere in the current except near the origin, and \eqref{logterm} must be used to capture the near-origin behaviour at very late times.

\subsection{Comparison with numerical simulations}

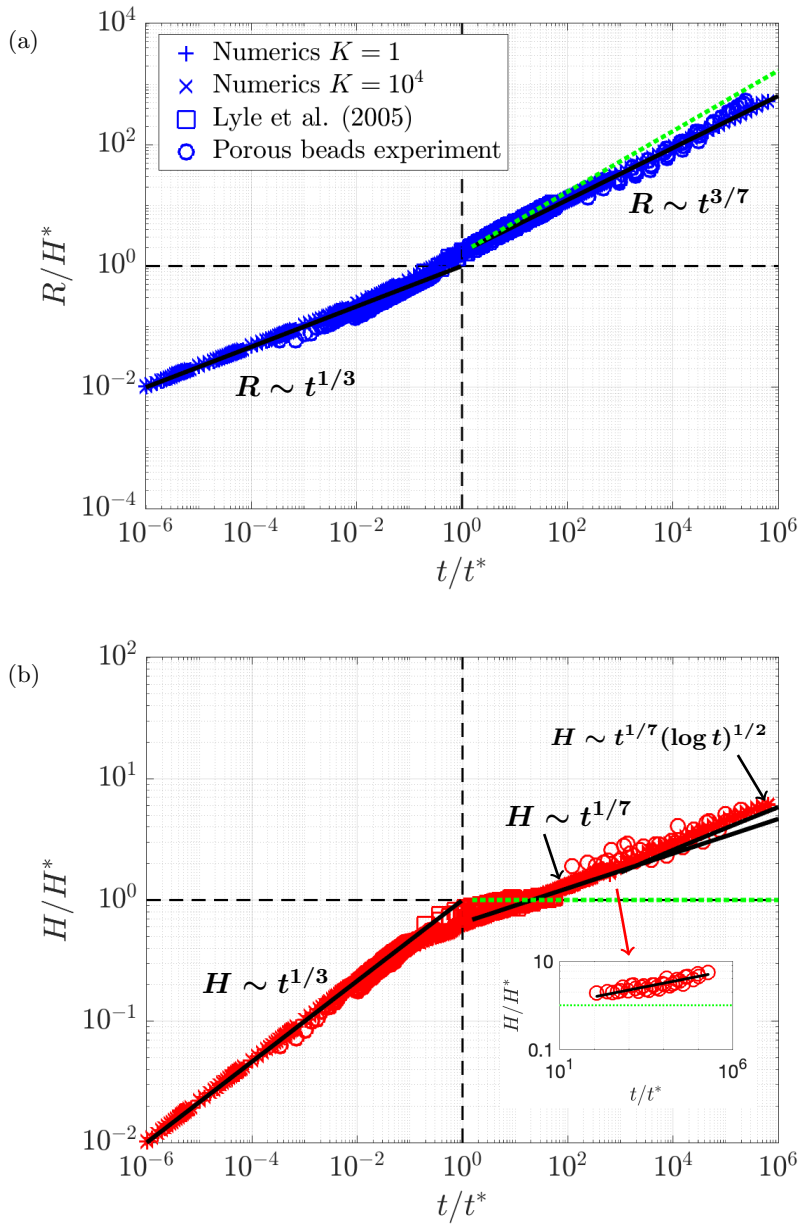
\begin{figure}
\centering
\begin{tikzpicture}[scale=0.8]
\node at (0,0) {\includegraphics[width=0.8\textwidth]{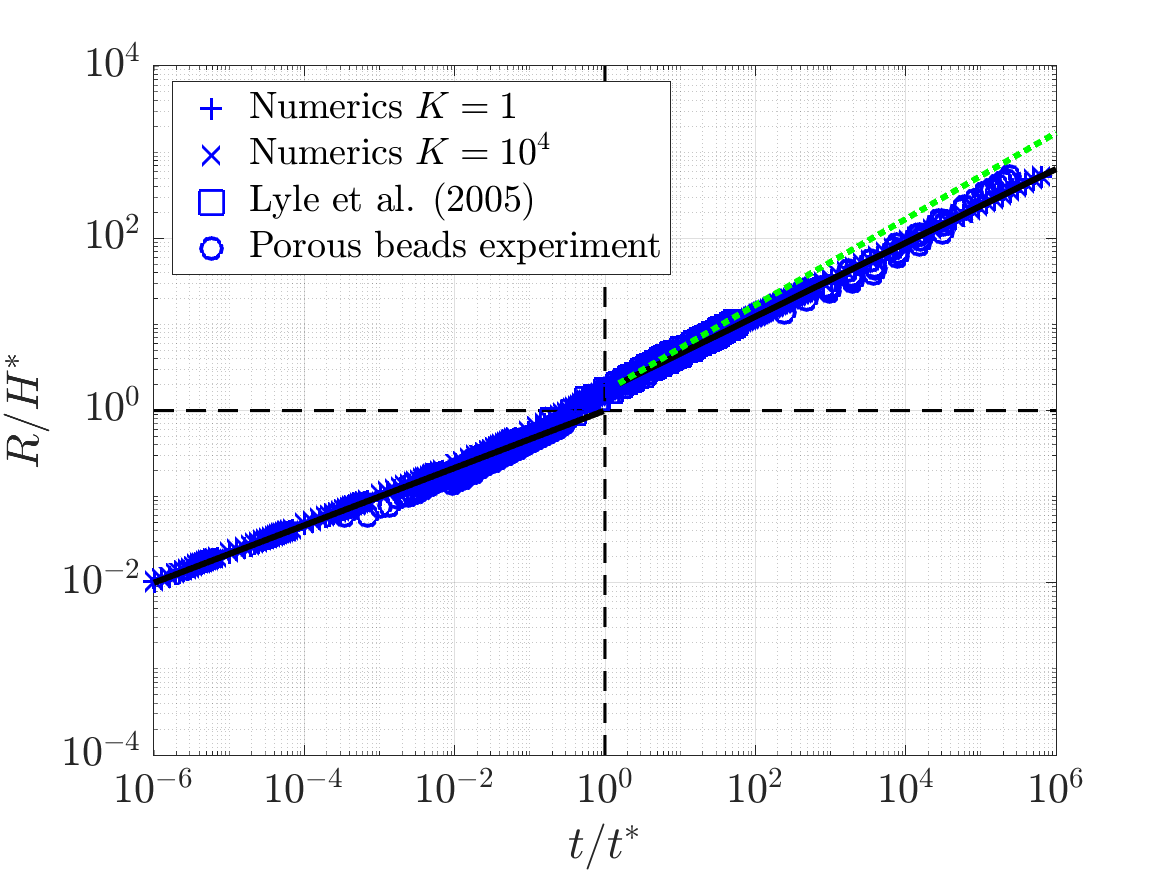}};
\node[black] at (4,1.4) {\large $\boldsymbol{R\sim t^{3/7}}$};
\node[black] at (-2.5,-1.6) {\large $\boldsymbol{R\sim t^{1/3}}$};
\node at (-7,4) {(a)};
\end{tikzpicture}
\begin{tikzpicture}[scale=0.8]
\node at (0,0) {\includegraphics[width=0.8\textwidth]{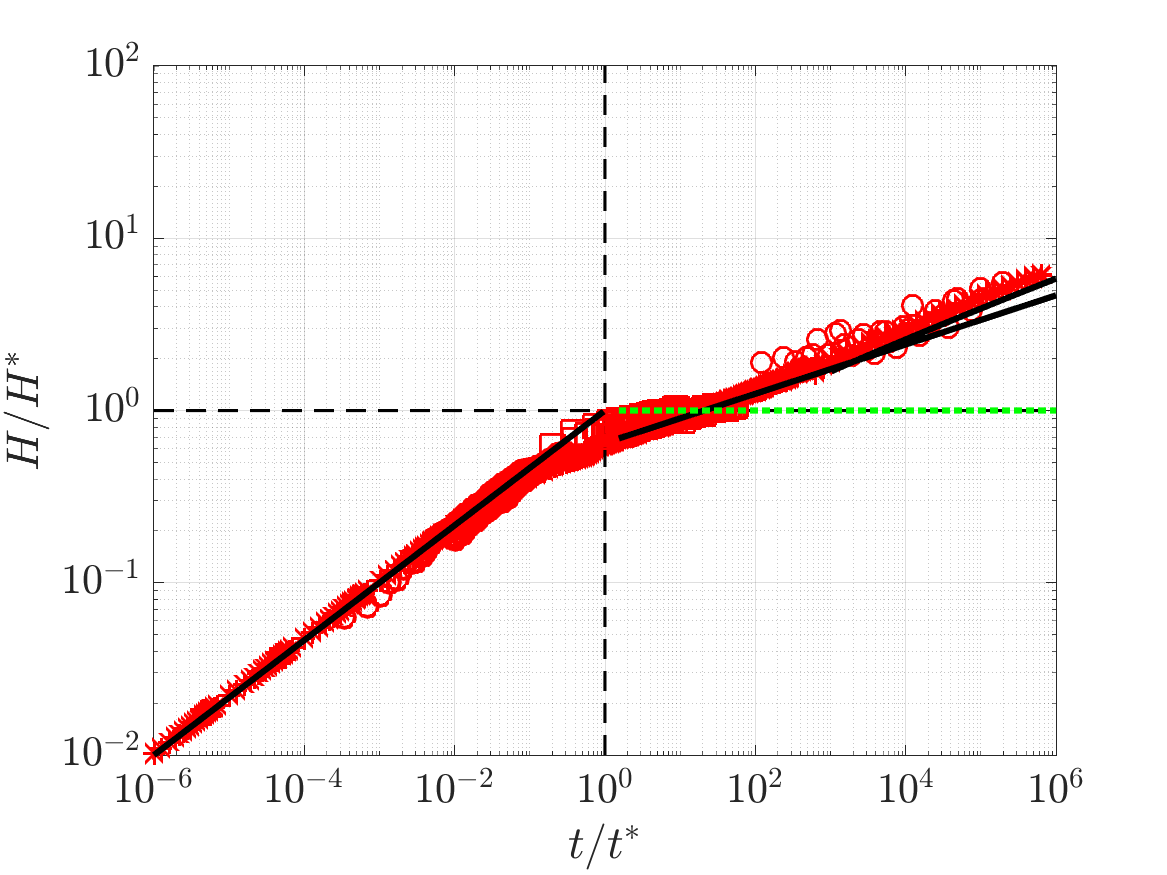}};
\node[black] at (3.5,3) { $\boldsymbol{H\sim t^{1/7}(\log t)^{1/2}}$};
\node[black] at (2.,1.75) {\large $\boldsymbol{H\sim t^{1/7}}$};
\node[black] at (-3,-1) {\large $\boldsymbol{H\sim t^{1/3}}$};
\node at (3,-1.8) {\includegraphics[width=0.25\textwidth]{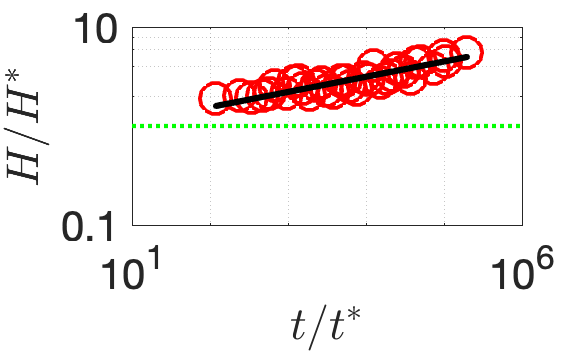}};
\draw[line width=1,->,red] (2.8,0.5) -- (3,-0.6) ;
\draw[line width=1,->] (4.8,2.7)  -- (5.25,1.95)  ;
\draw[line width=1,->] (1.6,1.4)  -- (1.85,0.65)  ;
\node at (-7,4) {(b)};
\end{tikzpicture}
\caption{Analytical, numerical and experimental data for the radial (a) and vertical (b) extents of the current, $R$, $H$. Numerical results are shown for both an isotropic medium $K=1$ and an anisotropic medium $K=10^4$, whereas experimental results are only for the isotropic case. Analytical early and late time scalings \eqref{early_2d_anis1}-\eqref{early_2d_anis2},\eqref{late_rad_isoH}-\eqref{late_rad_isoRp}, are shown with solid black lines and alternate scalings derived by \citet{lyle2005axisymmetric} ($R\sim t^{1/2},H\sim H^*$) are shown with dotted green lines. Best fit power law $H/H^*\propto (t/t^*)^{0.1537\pm0.0176}$ for the late time thickness data is shown as an insert in (b).
\label{converge2}}
\end{figure}

We compare our analytical predictions for the early-time, late-time, and transition behaviour, to finite difference numerical simulations of axisymmetric Darcy flow in an anisotropic porous medium. Before discussing these results, we first briefly outline the numerical method. We model the flow using the Darcy equations (in cylindrical polar coordinates), which are
\begin{align}
\mathbf{u}_i=-\frac{1}{\mu}\mathbf{k}\nabla \lb p + \rho_i g z \rb, \quad \nabla \cdot \mathbf{u}_i=0,\quad \mathrm{in} \quad \Omega_i,\quad i=1,2,\label{eqn1}
\end{align}
where $\rho=\rho_1$ inside the injected fluid domain (which we denote $\Omega_1$) and $\rho=\rho_2$ outside the injected fluid domain (which we denote $\Omega_2$). The anisotropic permeability matrix $\mathbf{k}$ is accounted for by imposing different values, $k_H$ and $k_V$, in the radial and vertical directions.

The numerical solution at $r=0$ is not available due to the singular nature of the divergence in the continuity equation $\nabla \cdot \mathbf{u}=0$. Instead, the domain begins at a small but finite value of $r_0=0.05H^*$ close to the origin (we note that the solution is insensitive to the precise value $0.05$).
The governing equations \eqref{eqn1} are accompanied by boundary conditions corresponding to impermeable/symmetric walls at $r=r_0$ and $z=0$, except for a constant input flux $Q$ injected over a small region near the origin, and far-field (Dirichlet) pressure conditions imposed at the boundaries of a large but finite domain.

The interface between the domains $\Omega_1$ and $\Omega_2$ is given by a set of points $\partial \Omega=\mathbf{r}(t)$, which is treated as a passive scalar and is therefore advected at the fluid velocity, such that
\beq
\frac{\mathrm{d}\mathbf{r}}{\mathrm{d}t}=\frac{1}{\phi}\mathbf{u}(\mathbf{r}),\label{dyneq}
\eeq
with initial conditions given by a small surface $\mathbf{r}(0)=\mathbf{r}_0$ close to the origin. 
Following the Lagrangian approach, the interface is discretised and grid points are advected along streamlines. Therefore, at every time-step the domains $\Omega_1$ and $\Omega_2$ are updated via interpolation over a gridded mesh of 150$\times$150 points. Likewise the grid is stretched over time to account for the fact that the fluid region grows across several orders of magnitude.
The dynamic equation for the fluid-fluid interface \eqref{dyneq} is solved using an explicit fourth order Runge-Kutta scheme with an adaptive time-step, where at each time-step the Darcy equations, \eqref{eqn1}, (with updated domains $\Omega_i$) are solved using a second order finite difference scheme. A small amount of Gaussian smoothing is applied to the interface at every time-step to aid stability, but this is done under the constraint of mass conservation.

In figure \ref{twod_stream} the gravity current shape $h(r,t)$ is shown together with streamlines at different times. 
To facilitate displaying the data on similar scales, we stretch the horizontal and vertical coordinates using the early and late scalings.
At early times $t\ll t^*$, the flow leaves the origin in straight lines, and the shape of the current is given by \eqref{isotropic}. 
Later, when $t\approx t^*$, the flow enters a transition regime in which gravity becomes appreciable in the majority of the current, except in the near-origin pressure-driven region, whose relative radial extent $R_p(t)/R(t)$ diminishes in size like $\sim t^{-2/7}$. Much later, when $t\gg t^*$, the shape of the gravity current converges to the approximate solution \eqref{approxsol}, and the streamlines away from the origin become horizontal. We also plot the horizontal and vertical extents $R$, $H$, in figure \ref{converge2}. At early ($t/t^*<10^{-2}$) and late ($t/t^*>10^{2}$) times, the numerical solution matches the analytical scalings \eqref{early_2d_anis1}-\eqref{early_2d_anis2},\eqref{late_rad_isoH}-\eqref{late_rad_isoR} closely. At very late times, the slow logarithmic growth \eqref{logterm} can be observed as a slight deviation in the vertical extent $H$ away from the late time scaling \eqref{late_rad_isoH}, which is consistent with our predictions.

To understand the role of the inner pressure-driven region, we have also calculated the maximum absolute value of the pressure deviation from hydrostatic $|p-p_\mathrm{hyd}|_\infty$, normalised by the weight of the injected fluid $\Delta \rho g H$. This deviation does not fall to zero over time, but instead appears to tend to a constant value of around $|p-p_\mathrm{hyd}|_\infty/\Delta \rho g H\approx 3$. This indicates that the flux is not well approximated by the gravity-driven component \eqref{gravfluxdef} everywhere within the current. As a result, the classical thin-film equation \eqref{thinfilm} and the similarity variables \eqref{simvars} cannot be accurate in the late-time regime. In particular, the boundary condition at the origin feeding the gravity current with flux $Q$ cannot be supplied with a gravity-driven term of the form \eqref{gravfluxdef} as this would cause an infinite thickness. Instead, the only way for the thickness to remain finite is with a pressure-driven input flux at the origin, which explains why the pressure deviation from hydrostatic may never drop to zero over time. 

We have also performed the same numerical calculation in the case of a two-dimensional injection (from a line-source) and in this case the pressure deviation $|p-p_\mathrm{hyd}|_\infty/\Delta \rho g H$ does drop to zero over time, indicating that (by contrast) the two-dimensional flux is well approximated by the corresponding gravity-driven component ($-u_b  h \partial h/\partial x$) everywhere within the current at late times.

Since the power laws $t^{3/7}$ and $t^{1/2}$ are not hugely different, we have also tried comparing the numerical solution with the classical self-similar scalings \eqref{simvars} proposed by \citet{lyle2005axisymmetric}, as shown with dotted green lines in figure \ref{converge2}. Since the disparity between the different proposed scalings grows larger over time like $\sim t^{1/2-3/7}\sim t^{1/14}$ (in the case of radius) and like $\sim t^{0-1/7}\sim t^{-1/7}$ (in the case of thickness), after a time scale $t/t^*=10^6$ this corresponds to disparaging factors of $10^{6/14}\approx 2.68$ and $10^{-6/7}\approx 0.14$, respectively. Since these discrepancies lie outside the 5$\%$ error observed in figure \ref{converge2}, this indicates that the scalings \eqref{late_rad_isoH}-\eqref{late_rad_isoR} are more accurate than the self-similar scalings \eqref{simvars}.

\subsection{Comparison with experimental data}
\label{sec_exp}

\begin{table}
\centering
\begin{tabular}{|c|c|c|c|c|c|c|c|c|}
\hline
$Q$ (cm$^3$/s)  &  0.020  &   0.021 &  0.043 &  0.125 &  0.209  & 17.36  & 17.67  & 31.62 \\
\hline
$g'$ (cm/s$^2$) & 172.66  & 178.54  &  159.90 & 159.90 & 151.07 & 22.56 & 22.56 & 22.56 \\
 \hline
\end{tabular}
\caption{List of experiments and corresponding parameter values. \label{tab_exp}}
\end{table}

Next we compare our results with experiments conducted in a porous media tank. 
We use the same apparatus for our experiments as \citet{lyle2005axisymmetric} except that we use a peristaltic pump instead of a fluid reservoir to control the flow rate (for improved accuracy). 
A square perspex tank with a  $61\un{cm}\times61\un{cm}$ base is filled with 0.3 cm glass Ballotini beads saturated with fresh water. 
Due to the difficulties associated with creating an anisotropic permeability field from these beads, our experiments are restricted to the isotropic case. Hence, the dependence on the anisotropy parameter $K$ in the analytical scalings \eqref{early_2d_anis1}-\eqref{early_2d_anis2},\eqref{late_rad_isoH}-\eqref{late_rad_isoRp} is checked against our numerical simulations, whilst the dependence on the remaining parameters (e.g. dimensional scalings and power laws) is checked against both numerical simulations and experiments (in the isotropic case).

Two cameras are positioned above and alongside the tank, pointing vertically downwards (plan view) and horizontally (side view), to capture both the radius and the thickness of the current. Salty water dyed with red food colouring is injected into the corner of the tank at different salt concentrations leading to a variety of density contrasts. By injecting into one corner of the tank the experiment represents one quarter of the full axisymmetric scenario whilst allowing for a larger range of radial data given the tank size. 

A full list of the different input flow rates and reduced gravity values $g'=\Delta \rho g/\rho_w$ (where $\rho_w=1\un{g/cm^3}$ is the density of water) is given in Table \ref{tab_exp}.
The experiments at high flow rates $Q=17.36-31.62\un{cm^3/s}$ result in a current with an order $\mathcal{O}(1)$ aspect ratio, with most of the injected fluid located in the interior region of the tank (away from the walls). Hence, the porosity and permeability are estimated using the values calculated for randomly packed beads, $\phi=0.37$ and $k=6.8 \times 10^{-9}\un{m}^2$ \citep{lyle2005axisymmetric}. However, the experiments conducted at low flow rates $Q=0.020-0.209\un{cm^3/s}$ result in a gravity current no thicker than 1-3 bead diameters ($z=d-3d$) and hence the flow is subject to wall effects. In these cases the porosity is estimated using the empirical formula derived by \citet{ribeiro2010mean} accounting for wall effects, which is
\beq
\phi=0.373+0.917e^{- 0.824 \,z/d }\approx0.55,\label{empiriphi}
\eeq
where we have used an average flow depth value $z=2d$, and the permeability is estimated using the Kozeny-Carman relationship
\beq
k=\frac{d^2\phi^3}{180(1-\phi)^2}\approx 4.36 \times10^{-8}\un{m}^2.
\eeq
It is worth noting how the shear in permeability due to wall effects \eqref{empiriphi} may modify the flow. As discussed by \citet{hinton2018buoyancy}, whilst such shear effects may alter the coefficients of the spreading rates, they do not modify the power laws, which motivates the use of different constant values here.

\begin{figure}
\centering
\begin{tikzpicture}[scale=0.8]
\node at (0,0) {\includegraphics[width=0.45\textwidth]{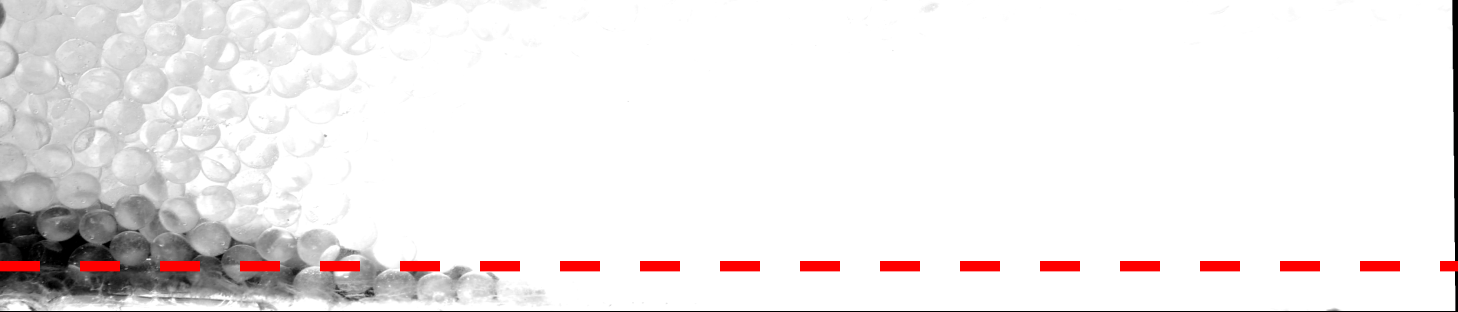}};
\node at (0,-2) {\includegraphics[width=0.45\textwidth]{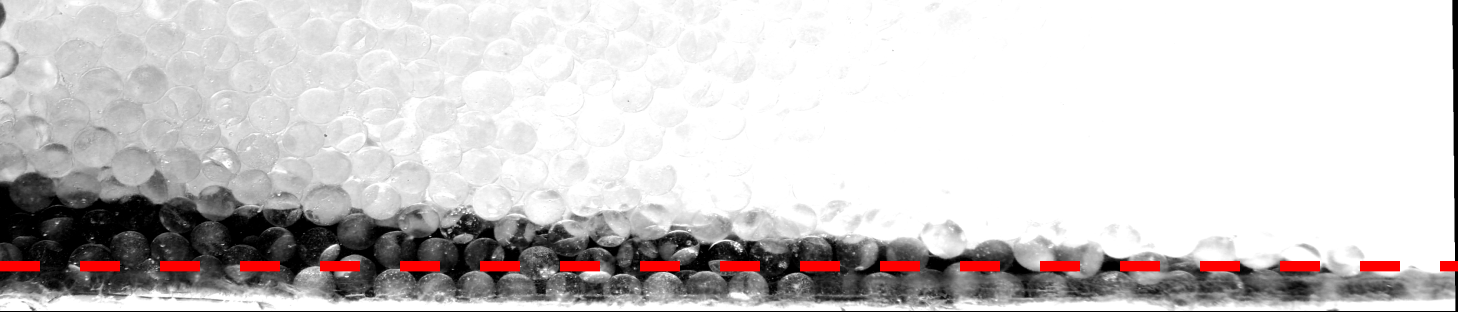}};
\node at (0,-4) {\includegraphics[width=0.45\textwidth]{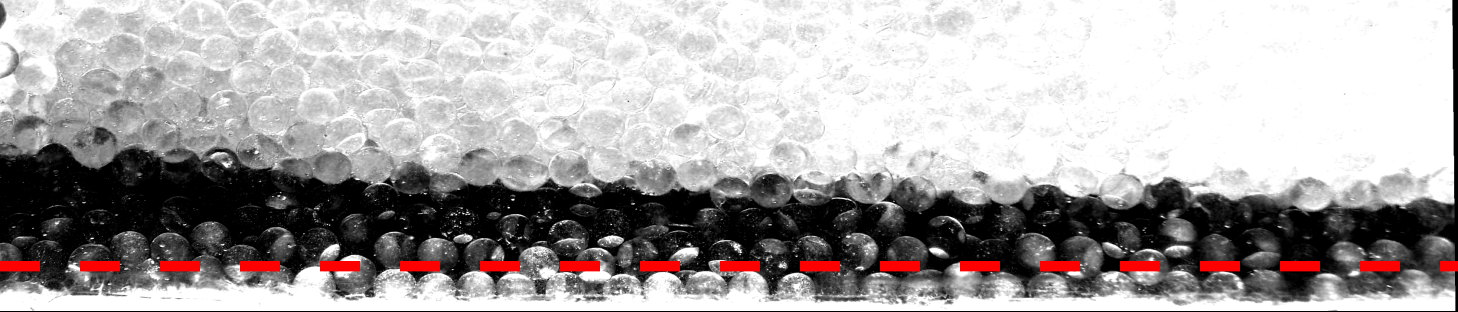}};
\node at (8,0) {\includegraphics[width=0.45\textwidth]{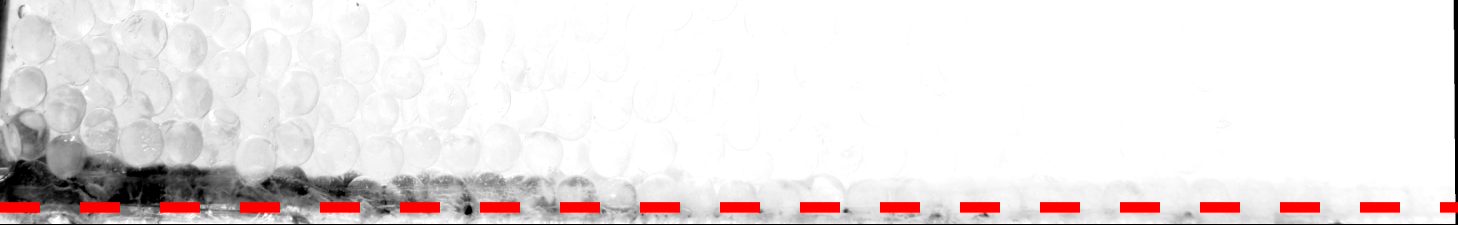}};
\node at (8,-2) {\includegraphics[width=0.45\textwidth]{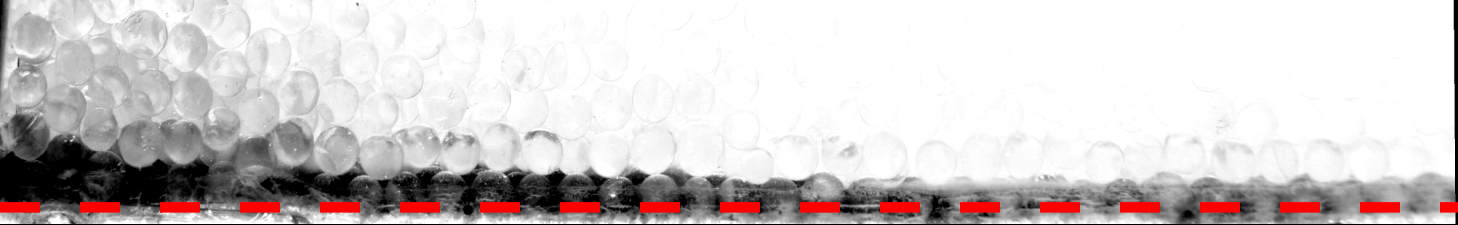}};
\node at (8,-4) {\includegraphics[width=0.45\textwidth]{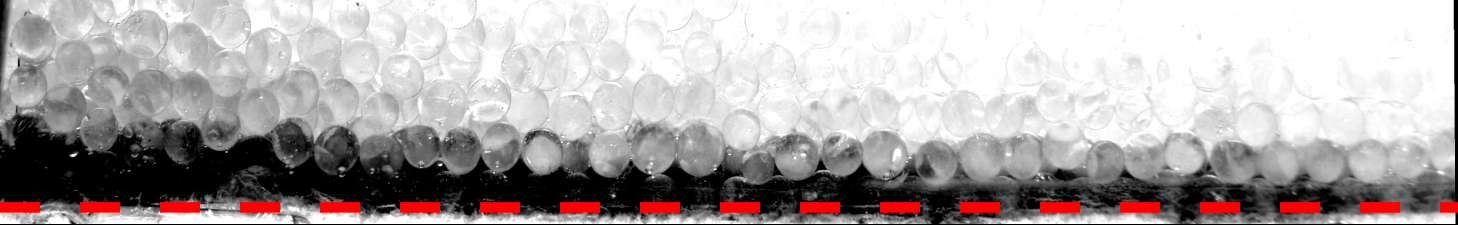}};
\node[red] at (-0.8,-0.2) {$\boldsymbol{H=H^*}$};
\node at (2,0.5) {$\boldsymbol{t/t^*=119}$};
\node at (2,-1.5) {$\boldsymbol{t/t^*=1908}$};
\node at (2,-3.5) {$\boldsymbol{t/t^*=30532}$};
\node at (10,0.3) {$\boldsymbol{t/t^*=1540}$};
\node at (10,-1.7) {$\boldsymbol{t/t^*=12316}$};
\node at (10,-3.7) {$\boldsymbol{t/t^*=197060}$};
\node at (0,1.5) {\bf $\boldsymbol{Q=0.209$ cm$^3$/s, $H^*=0.36\un{cm}}$};
\node at (8,1.5) {\bf $\boldsymbol{Q=0.020$ cm$^3$/s, $H^*=0.11\un{cm}}$};
\node at (-4,1.5) {(a)};
\node at (4,1.5) {(b)};
\end{tikzpicture}
\caption{(a,b) Closeup snapshots of the gravity current thickness near the origin during two injection experiments with different flow rates. Photos are taken over 30 minutes (a) and 1 hour (b), and spaced logarithmically in time.
\label{expt3}}
\end{figure}
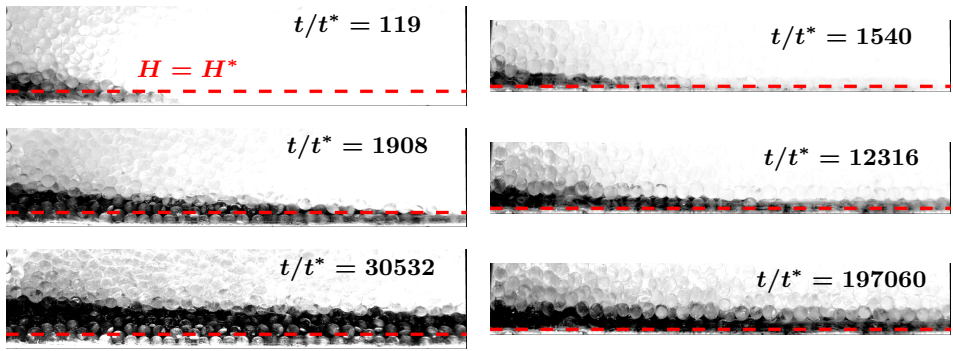

By varying the flow rate and reduced gravity, a variety of different transition time scalings are achieved, producing data in the range $t/t^*=10^{-3}-10^{5}$. We also complement our data with the experiments of \citet{lyle2005axisymmetric} which are in the intermediate range $t/t^*=0.08-22$. To reach such late time values we inject using a very slow flow rate of $Q=0.02\un{cm^3/s}$ over one hour with photos taken at logarithmically spaced time intervals starting at 30 s (see figure \ref{expt3}). After post-processing the images, the radial and vertical extents of the current are extracted by dividing the pixels into binary values according to a threshold value (enabled by the colour contrast due to the dye). 

Data for the thickness and radius of the current are plotted together with numerical and analytical results in figure \ref{converge2}. The radius data $R(t)$ approximately follows the $3/7$ power law \eqref{late_rad_isoR}, however one could argue that the $1/2$ power law \eqref{simvars} is just as accurate within the error margins of the experimental data. More clarity is achieved  by observing the variation of the thickness near the origin $H(t)$, as shown by the photos in figure \ref{expt3}. In particular, the thickness is clearly observed to increase beyond the transition value by as much as $H\approx  5.5 H^*$ over the course of the experiments, albeit with a very small growth rate. Indeed, without running the experiments for such a long time it would appear that the thickness has ceased growing entirely. By contrast, when observed at logarithmically spaced time intervals the thickness shows no signs of abating. Post-processing of the images to extract the interface position reveals that the thickness follows the 1/7 power law \eqref{late_rad_isoH} to good approximation, as shown in figure \ref{converge2}. We have also tried using a least-squares minimisation that fits an arbitrary power law $H= A t^B$ to the combined experimental dataset (over four different flow rates) at late times $t>100 t^*$, as shown in the figure insert. This produces a power law of $B=0.1537\pm 0.0176$ which is close to $1/7\approx 0.1429$. Moreover, this slightly larger numerical value for the power law might hint towards the faster logarithmic growth predicted earlier. 

It is not possible to discern the logarithmic correction to the thickness \eqref{logterm} quantitatively since this would require much longer time scales than we could achieve experimentally. Nor is it possible to accurately discern the convergence of the shape of the current to the profile \eqref{approxsol} over long times, since the aspect ratio of the current is incredibly slender (around 200:1). To extend the experiments to larger values of $t/t^*$ is challenging in practice, since smaller flow rates than $Q=0.02\un{cm^3/s}$ would result in current thicknesses much smaller than a single bead size ($H^*\ll 0.3\un{cm}$) that would be difficult to observe. Alternately one could run the experiments for a longer time, but this requires a much larger tank than we have available. For example to extend $t/t^*$ by a factor 100 would require a tank of dimensions $100^{3/7}\approx 7.2$ times larger (i.e. with a $439\un{cm}\times 439\un{cm}$ base).
Nevertheless, our experiments clearly show an increasing thickness near the origin with the approximate 1/7 power law behaviour \eqref{late_rad_isoH}. Hence, by conservation of mass \eqref{massconlate} it follows that the radius must grow according to a 3/7 power law, and the shape profile \eqref{approxsol} must become a good approximation over long times (e.g. see analysis in Section \ref{sec_seventh}).


It is important to note the possible effects of diffusion/dispersion over such long time scales. The length scales for molecular diffusion, fluid dispersion, and Taylor dispersion are approximately $\ell \sim (D t)^{1/2}$, $\ell \sim (d U t)^{1/2}$ and $\ell \sim (d^2 U^2  t / 48D)^{1/2}$, respectively, where $D$ is the molecular diffusivity of salt or dye (both similar values), $U$ is a characteristic velocity scale, and $d=0.3\un{cm}$ is the diameter of the Ballotini beads. As an estimate for the diffusivity we take $D=10^{-10}\un{m^2/s}$, and for the velocity we take the value at the top of the current $U\sim \mathrm{d}H/\mathrm{d}t$. Inputting these, we calculate diffusion/dispersion scales of $\ell \sim 0.06 \un{cm}$, $\ell \sim 0.13 \un{cm}$ and $\ell \sim 0.04 \un{cm}$, which are all considered small within the experimental context. It is clear from the images in figure \ref{expt3} that the interface near the origin remains relatively sharp, indicating that the effects of dispersion are negligible there. Hence, this validates our approach of post-processing the images to extract the current thickness at late times.

\section{Relevance to carbon storage sites}
\label{sec_ccs}

In this section we situate our results within the context of carbon storage applications, in which CO$_2$ is injected into a porous reservoir saturated with brine and bounded above by an impermeable cap rock (note that under the Boussinesq approximation our analysis equally applies to a lighter fluid injected into a heavier fluid which is bounded above). In particular, the ability to determine whether an injected CO$_2$ plume is in a pressure-driven regime or a gravity-driven regime is useful for determining the expected shape, and the expected plume migration speeds. This is important for ensuring that the CO$_2$ can be stored as safely and efficiently as possible.

For this analysis we compare our estimates for the transition time and thickness scalings $t^*$, $H^*$, to typical parameter values from field sites. This enables us to estimate whether the required injection time, or whether the confines of the aquifer (e.g. the available space between impermeable cap rocks) are sufficient for a gravity current to form. To make this comparison we require approximate parameter ranges for the injection flux $Q$, as well as the buoyancy velocity $u_b=k_H\Delta\rho g/\mu$, porosity $\phi$ and anisotropy $K=k_H/k_V$.

\begin{table}
\centering
{\large
\begin{tabular}{c|c|c|c|c|c|c}
 & Regime & Variable & Prefactor & Scaling &  $t^n$ &  $K^n$ \\
\hline
 & $t\ll t_x^*$ & $R_x$ & $2/\pi^{1/2}$ & $(Q_x/\phi)^{1/2}$ & 1/2 & 1/4 \\
 & $t\ll t_x^*$ & $H_x$ & $2/\pi^{1/2}$ & $(Q_x/\phi)^{1/2}$ & 1/2 & $-1/4$ \\
Planar & $t\gg t_x^*$ & $R_x$ & $\eta_N\approx 1.482$ & $(Q_x u_b/\phi^2)^{1/3}$ & 2/3 & $\sim$ \\
 & $t\gg t_x^*$ & $H_x$ & $f_0\approx 1.296$ & $(Q_x^2/u_b\phi)^{1/3}$ & 1/3 & $\sim$ \\
  & & $t_x^*$ & $\pi/4$ & $\phi Q_x/u_b^2 $ & $\sim$ & $3/2$ \\
    & & $H_x^*$ & $1$ & $Q_x/u_b $ & $\sim$ & $1/2$ \\
 \hline
 & $t\ll t^*$ & $R$ & $(3/2\pi)^{1/3}$ & $(Q/\phi)^{1/3}$ & 1/3 & 1/6 \\
 & $t\ll t^*$ & $H$ & $(3/2\pi)^{1/3}$ & $(Q/\phi)^{1/3}$ & 1/3 & -1/3 \\
 Radial & $t\gg t^*$ & $R$ & $\zeta_N=( 2^{11/2}/\pi^2)^{1/7}$ & $(Q^2 u_b/\phi^3)^{1/7}$ & 3/7 & 1/14 \\
 & $t\gg t^*$ & $H$ & $\xi_0=(2^{1/2}\pi^3)^{-1/7}$ & $(Q^3/u_b^2\phi )^{1/7}$ & 1/7 & $-1/7$ \\
   & & $t^*$ & $2\pi/3$ & $\phi (Q/u_b^3)^{1/2} $ & $\sim$ & $1$ \\
   & & $H^*$ & $1$ & $(Q/u_b)^{1/2} $ & $\sim$ & $0$ \\
 \hline
\end{tabular}
}
\caption{List of asymptotic limiting behaviours for the horizontal and vertical extent of the flow, as well as the transition scalings, for the case of planar injection (from a line source) and radial injection (from a point source). For example, the late time extent in the radial case is given by $R=\zeta_N(Q^2 u_b/\phi^3)^{1/7}t^{3/7}K^{1/14}$. Note, the vertical extent in the case of radial injection $H$ has a slow logarithmic dependence \eqref{logterm} which we emit here for simplicity. \label{table1}}
\end{table}

As described by \citet{huppert2022fate} in the isotropic case, the time to transition to a gravity current may be significantly prolonged by small permeability values. Here we show that the transition time \eqref{transtimescale} may be further prolonged by the presence of anisotropy, such that even sedimentary systems with large horizontal permeability values $k_H$ may still remain in a pressure-driven regime for a long time (e.g. hundreds or thousands of years) if the vertical permeability $k_V$ is significantly smaller ($K\gg1$). 
This has significant consequences for modelling approaches, which often directly assume a gravity-driven plume within the timeframe of the injection site, and indicates the need for detailed measurements of heterogeneities to ensure accurate predictions for the migration of CO$_2$.

In addition to the case of axisymmetric radial injection, which is the focus of the current study, we also briefly describe the analogous case of two-dimensional planar injection (from a line-source) into anisotropic media. This is a simple extension of the study by \citet{huppert2022fate} for isotropic media, incorporating different permeabilities $k_H$, $k_V$, in the horizontal and vertical directions. As such, for the present analysis we skip the derivation of these scalings and simply present them in Table \ref{table1} for reference. 
To distinguish the planar case from the radial (axisymmetric) case, we introduce subscript notation for the transition time $t^*_x$, the transition thickness $H^*_x$, and the injection flux per unit width $Q_x$. Likewise, the variables $R_x$ and $H_x$ in the planar case are equivalent to the lateral and vertical extents of the current.
We note that the dynamics of the current for planar injection only depend on the anisotropy parameter $K$ at early times, not at late times, as discussed in the Introduction.

The references for the parameter values used comprise a variety of sources describing different carbon storage sites around the world \citep{bickle2007modelling,chadwick2009flow,oldenburg2011leakage,golding2013effects,cowton2016inverse,williams2017improved,bickle2017rapid,cowton2018benchmarking}.
The buoyancy velocity is estimated using parameter values $\mu= \left[5.5-6.6\right]\times 10^{-5}\un{Pa}\cdot\un{s}$, $\Delta\rho=232-309\un{kg/m}^3$, and $k_H=10^{-14}-10^{-12}\un{m}^2$, resulting in $u_b=\left[3.4-551\right]\times 10^{-7}\un{m/s}$. For the injection flux, we take estimates for a two-dimensional line source as $Q_x=\left[0.4-3\right]\times 10^{-4}\un{m}^2/\un{s}$, and for a radial point source as $Q=\left[0.1-4\right]\times 10^{-2}\un{m}^3/\un{s}$. Porosity is taken as $\phi=0.2-0.25$. 
The degree of anisotropy varies considerably between CO$_2$ sequestration sites. For example, observations of permeability variations were around two orders of magnitude for Salt Creek, USA \citep{bickle2017rapid}, and four orders of magnitude for the Tilje formation in Norway (a potential sequestration site) \citep{martinius1999multi,bergmo2017quality}. Hence, for this study we consider anisotropy in the range $K=1-10^4$.

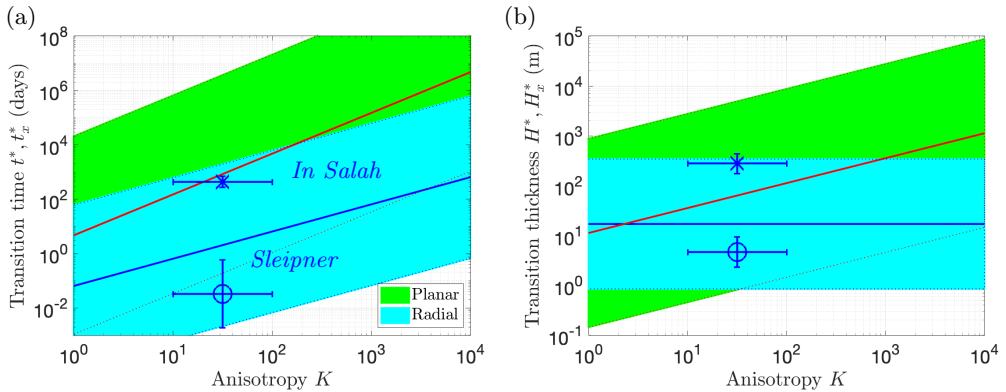
\begin{figure}
\centering
\begin{tikzpicture}[scale=0.8]
\node at (0,0) {\includegraphics[width=0.5\textwidth]{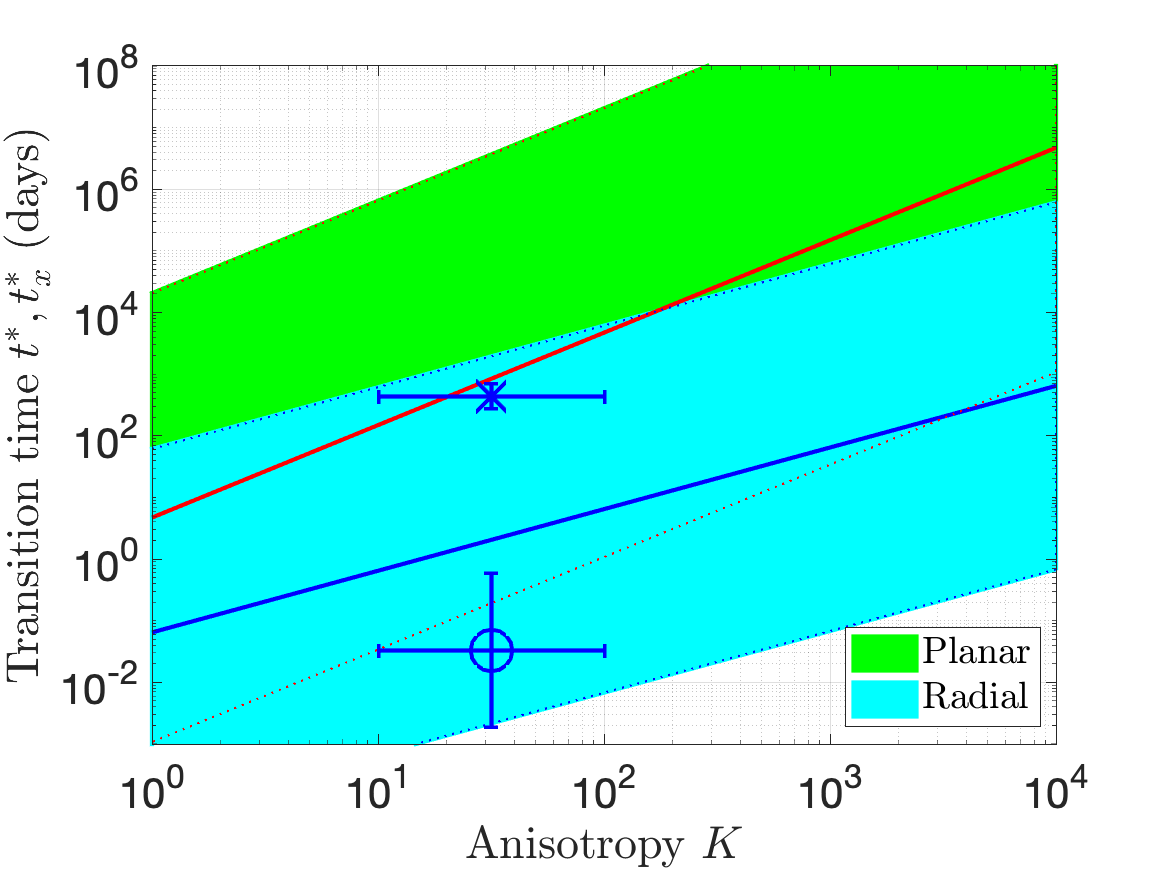}};
\node at (8.5,0) {\includegraphics[width=0.5\textwidth]{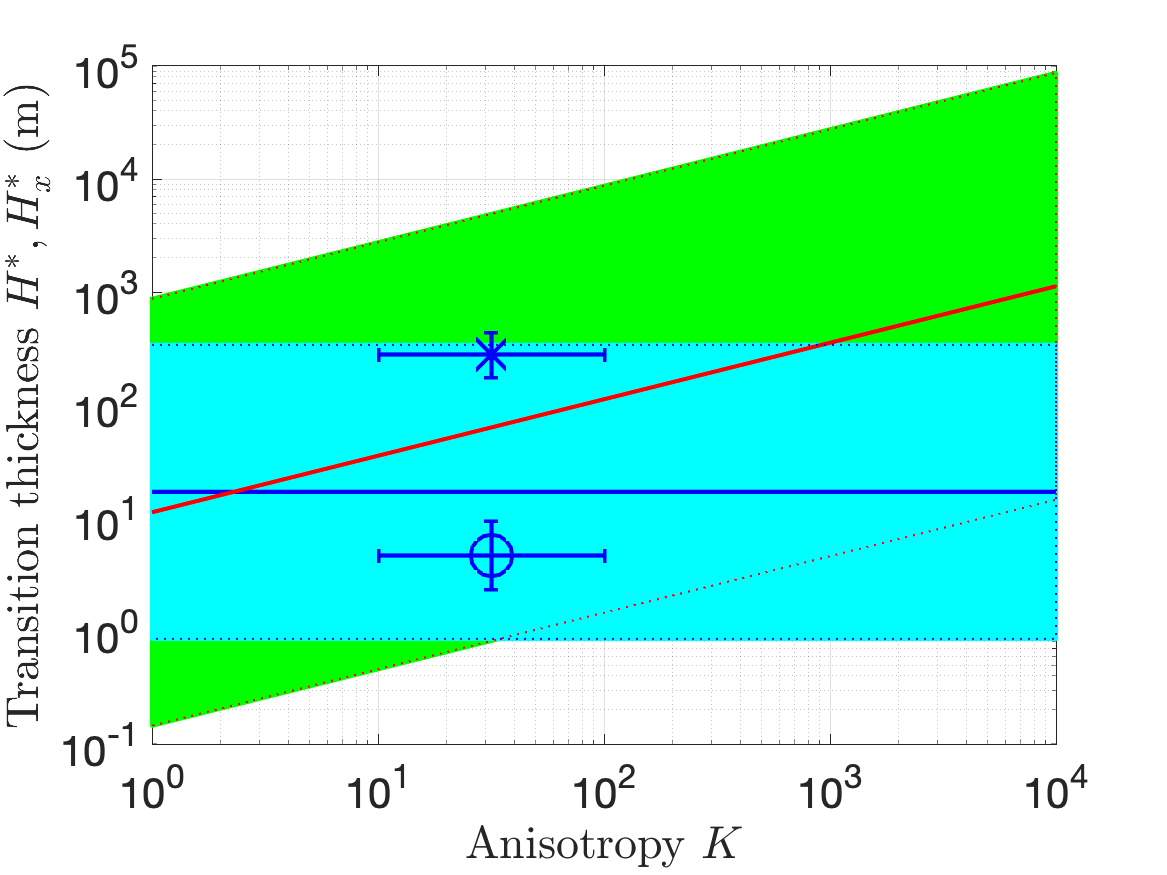}};
\node at (-4,3) {(a)};
\node at (4.25,3) {(b)};
\node[blue] at (1.2,0.5) {\it In Salah};
\node[blue] at (0.5,-1) { \it Sleipner};
\end{tikzpicture}
\caption{Transition time $t^*$, $t_x^*$ (a) and transition thickness $H^*$, $H_x^*$ (b) for different anisotropy values $K$, in the case of planar (line-source) injection and radial (point-source) injection. Corresponding data for the Sleipner and In Salah carbon sequestration sites (assuming radial injection) are illustrated, together with uncertainty estimates for parameter values. \label{transition}}
\end{figure}

Collating all of this data, we plot the transition time $t^*$ and transition thickness $H^*$ in figure \ref{transition} for a range of anisotropy values, in both the planar and radial cases. Upper and lower bounds are displayed which correspond to the range of possible values for the injection flow rates $Q_x$, $Q$,  buoyancy velocity $u_b$, and porosity $\phi$.
For comparison, we also display data points for two CO$_2$ sequestration sites: Sleipner, Norwegian North Sea (in operation 1996 - present), and In Salah, Algeria (in operation 2004 - 2012).
Taking both of these as point source (radial) injection sites, and using approximate anisotropy values $K=10-100$, we calculate the transition time scale as less than a day for Sleipner and 280-700 days for In Salah. 
Likewise the transition thickness for Sleipner is around 2-9 m, as opposed to 170-440 m for In Salah.
This indicates that the modern day plume at Sleipner is very likely to be in a gravity-driven regime, whereas the CO$_2$ flow at the less permeable site at In Salah is  likely to have been in a pressure-driven regime for a significant period of operation. In particular, the large transition thickness at In Salah indicates that interaction with other confining stratigraphy (i.e. within 440 m of the injection point) was a likelier dominant control on CO$_2$ migration rather than gravity-driven spreading.

\section{Discussion}

The transition from a pressure- to a gravity-driven flow has been addressed for constant axisymmetric injection into an anisotropic porous medium. We have derived scalings for the early and late time growth of the current, as well as the timescale at which the transition occurs, showing close comparison with analogue experiments and numerical simulations. 
Early time scalings revealed that anisotropy causes a long and thin current in which the pressure is not hydrostatic, with a delayed transition to the gravity-driven regime. 
Hence this study presents a paradigm shift for such flows, which are typically assumed to be gravity-driven if long and thin. 
Analysis of the late time regime showed that a region near the origin must remain pressure-driven.
Within this pressure-driven region, strong vertical velocities are required to deliver a uniform horizontal flow to the rest of the gravity-driven current. 
The pressure within this region cannot become hydrostatic, since this causes an unphysical singular current thickness near the origin. 
Therefore, the inner pressure-driven region serves to counteract this singularity, thereby providing a significant non-hydrostatic contribution to the flux even at late times. 

One consequence of this pressure component is a finite but ever-growing thickness near the origin (growing like $\sim (\log t)^{1/2}t^{1/7}$), which contrasts previous theories that assumed a constant thickness scaling.
Another consequence is that the late time growth of the gravity current remains affected by the vertical permeability value $k_V$ (and hence the anisotropy ratio $K$) due to vertical velocities near the origin.
This is in contrast to the two-dimensional (planar) case, in which anisotropy only affects the time to transition, not the late time behaviour.
Our theoretical predictions were confirmed by comparison with numerical simulations of anisotropic Darcy flow run across 12 orders of magnitude in time, and porous media tank experiments in the isotropic case.
Transition timescales were calculated for anisotropy values of carbon sequestration applications, indicating that some field sites may never reach the gravity-driven regime over the course of operation.

It should be noted that for some CO$_2$ sequestration sites injection occurs over a vertical interval of around $1-10\un{m}$, such that the flow is delivered uniformly across the current depth at source. In such cases the vertical velocities associated with the inner pressure-driven region described in this study will not emerge unless the thickness scaling $H^*=(Q/u_b)^{1/2}$ is larger than this interval. Consequently, in such cases the classical dynamics of self-similarity with a radial extent growing like $R\sim t^{1/2}$ described by \citet{lyle2005axisymmetric} remain accurate. Hence, the results presented here are only relevant for point source injection sites, or for sites with a vertical injection interval smaller than the transition thickness $H^*$ (as is the case for anisotropic aquifers).

Whilst we have mentioned carbon sequestration applications throughout this study, we have neglected several physical effects which are relevant to the flow of CO$_2$ in brine. For example, we have neglected the viscosity contrast between these phases for simplicity (CO$_{2}$ is typically around 20-30 times less viscous than brine). By extending this study to account for such effects, it is expected that the viscosity ratio $M=\mu_1/\mu_2$ will appear in the early and late dynamics for $H$ and $R$ wherever the anisotropy parameter $K$ does. This is because the viscosity ratio can only affect the dynamics whenever the flow of injected and ambient fluids are coupled together (just like $K$). It would also be interesting to extend these results to account for multiphase effects, similarly to \citet{golding2013effects} in the case of homogeneous media, and similarly to \citet{benham2021two} in the case of heterogeneous media.

There are several interesting perspectives from this work that are worth discussing. Firstly, the different dynamics between early and late times present an opportunity for inverse modelling to help interpret petrophysical information about the reservoir. For example, if it can be shown that the radial extent of the flow is growing according to the early time power law behaviour, then this can place bounds on the permeability and anisotropy of the flow due to heterogeneities. This is useful for field sites since information about the heterogeneities is often restricted to sparse core measurements or coarse seismic surveys.

It is also worth noting the analogy between the current study and the canonical case of a classical viscous gravity current. It is easy to follow the above analysis for the case of constant injection of a viscous fluid onto an impermeable substrate in either the planar or radial case. It transpires that similar scalings can be derived for the time and thickness required to transition from a pressure- to a gravity-driven flow. For example, in the radial case these are given by $H^*=(Q\mu/\Delta\rho g)^{1/4}$ and $t^*=(2\pi/3)(\mu^3/(\Delta \rho g)^3 Q)^{1/4}$. 
However, it is not clear how the inner pressure-driven region affects the viscous gravity current at late times. In particular, the no-slip condition at the base of the current produces a different flow pattern to that considered here, making it difficult to translate our other results. Nevertheless, one could investigate such flow patterns by introducing tracers at the source of the gravity current, shedding light on the size and role of the pressure-driven region at late times.

\acknowledgements{
This research is funded in part by the GeoCquest consortium, a BHP-supported collaborative project between Cambridge, Stanford and Melbourne Universities, and by a NERC consortium grant ``Migration of CO$_2$ through North Sea Geological Carbon Storage Sites'' (grant no. NE/N016084/1).
}\\

Declaration of Interests. The authors report no conflict of interest.\\

\bibliographystyle{jfm}
\bibliography{bibfile.bib}

\end{document}